\documentclass[twocolumn]{aastex631}
\renewcommand{\today}{12 July 2025}

\usepackage{amsmath}

\shorttitle{2021 Outburst of YY Herculis: Optical Follow-Up}
\shortauthors{L. S. Sonith and U. S. Kamath}
\begin{document}

% \title{The 2021 Hot-Type Symbiotic Outburst of YY Herculis: An Optical Follow-Up Study}

\title{The 2021 Hot-type Symbiotic Outburst of YY Herculis: An Optical Follow-up Study}

\author[0000-0002-2033-3051]{L. S. Sonith}
\affiliation{Indian Institute of Astrophysics, Koramangala, 560034, Bangalore, India }
\affiliation{Pondicherry University, R.V. Nagar, Kalapet, 605014, Puducherry, India.}

\author[0000-0003-1729-0190]{U. S. Kamath}
\affiliation{Indian Institute of Astrophysics, Koramangala, 560034, Bangalore, India }
\affiliation{Pondicherry University, R.V. Nagar, Kalapet, 605014, Puducherry, India.}

%% Note that the \and command from previous versions of AASTeX is now
%% depreciated in this version as it is no longer necessary. AASTeX 
%% automatically takes care of all commas and "and"s between authors names.

%% AASTeX 6.31 has the new \collaboration and \nocollaboration commands to
%% provide the collaboration status of a group of authors. These commands 
%% can be used either before or after the list of corresponding authors. The
%% argument for \collaboration is the collaboration identifier. Authors are
%% encouraged to surround collaboration identifiers with ()s. The 
%% \nocollaboration command takes no argument and exists to indicate that
%% the nearby authors are not part of surrounding collaborations.

%% Mark off the abstract in the ``abstract'' environment. 
\begin{abstract}
We have followed up on the hot-type classical symbiotic outburst reported in YY Her using the Himalayan Chandra telescope. The outburst coincides with the secondary minima of the system.
Approximately 12 similar brightening events have been reported between 1890 and 2020, with only the 1993 outburst being studied spectroscopically.
In our study, we monitored the system from 2021 to 2023, covering $\sim$1.5 orbital cycles, providing an opportunity to understand the spectral evolution of the outburst over a complete orbital period of YY Her. 
We found that the temperature and luminosity estimations based on emission line fluxes exhibit orbital phase dependence. The values estimated at phase 0.5, corresponding to the secondary minimum, were the most reliable.
The temperature of the hot component is \( \approx 1.41 \times 10^5 \, \text{K} \), and the luminosity is \( \approx 1020 \, L_{\odot} \) during the outburst, reduced to \( \approx 1.3 \times 10^5 \, \text{K} \) and \( \approx 830 \, L_{\odot} \) after one orbital cycle at phase 0.5.
Temperature estimations during the outbursts suggest that YY Her exhibits both hot-type (2021) and cool-type (1993) behavior, similar to another symbiotic star, AG Dra.
Using variations of the Ca II absorption lines, we confirmed the contribution of the ellipsoidal effect in secondary minima in the YY Her light curve.

% This example manuscript is intended to serve as a tutorial and template for
% authors to use when writing their own AAS Journal articles. The manuscript
% includes a history of \aastex\ and includes figure and table examples to illustrate these features. Information on features not explicitly mentioned in the article can be viewed in the manuscript comments or more extensive online
% documentation. Authors are welcome replace the text, tables, figures, and
% bibliography with their own and submit the resulting manuscript to the AAS
% Journals peer review system. The first lesson in the tutorial is to remind
% authors that the AAS Journals, the Astrophysical Journal (ApJ), the
% Astrophysical Journal Letters (ApJL), the Astronomical Journal (AJ), and
% the Planetary Science Journal (PSJ) all have a 250 word limit for the 
% abstract\footnote{Abstracts for Research Notes of the American Astronomical 
% Society (RNAAS) are limited to 150 words}. If you exceed this length the
% Editorial office will ask you to shorten it. This abstract has 161 words.
\end{abstract}

%% Keywords should appear after the \end{abstract} command. 
%% The AAS Journals now uses Unified Astronomy Thesaurus concepts:
%% https://astrothesaurus.org
%% You will be asked to selected these concepts during the submission process
%% but this old "keyword" functionality is maintained in case authors want
%% to include these concepts in their preprints.
\keywords{Symbiotic binary stars(1674) --- Spectroscopy (1558) --- Z Andromedae stars(1835) --- Spectroscopic binary stars(1557)}
%% From the front matter, we move on to the body of the paper.
%% Sections are demarcated by \section and \subsection, respectively.
%% Observe the use of the LaTeX \label
%% command after the \subsection to give a symbolic KEY to the
%% subsection for cross-referencing in a \ref command.
%% You can use LaTeX's \ref and \label commands to keep track of
%% cross-references to sections, equations, tables, and figures.
%% That way, if you change the order of any elements, LaTeX will
%% automatically renumber them.
%%
%% We recommend that authors also use the natbib \citep
%% and \citet commands to identify citations. The citations are
%% tied to the reference list via symbolic KEYs. The KEY corresponds
%% to the KEY in the \bibitem in the reference list below. 

\section{Introduction} \label{sec:intro}

Symbiotic stars are long-period interacting binaries consisting of a 
mass-losing giant of a late spectral type and a hot component, typically a white dwarf (WD), surrounded by a circumstellar nebula (see the reviews by \citealp{2012BaltA..21....5M} \& \citealp{2019arXiv190901389M}). Symbiotic stars exhibit variability as a result of orbital motion and occasional outbursts. Three different types of outbursts are reported in symbiotic stars: symbiotic novae or slow novae, symbiotic recurrent novae, and classical symbiotic outbursts (Z And-type).
Symbiotic novae and symbiotic recurrent novae are rare events, since they require prolonged mass accretion onto the surface of WD before a thermonuclear runaway can be triggered.
In contrast, classical symbiotic outbursts are commonly seen outburst in symbiotic stars, typically showing 1--3 B mag brightening. These outbursts are either caused by the release of potential energy from extra-accreted matter or by a shift in the emission of hot component towards longer wavelengths, resulting from a radius expansion followed by an increased mass accretion rate \citep{2019arXiv190901389M}.
A detailed study by \cite{AG_DRA_Gonzalez_1999} on the classical symbiotic outbursts of AG Dra using the International Ultraviolet Explorer (IUE) combined with optical observations showed that they can be further classified into hot and cool types based on the temperature of the hot component during the outburst relative to its quiescent temperature.
During 2007-09 outbursts of AG Dra, \cite{2009PASP..121.1070M} noted that the He II to H$\beta$ ratio is enhanced during the hot type outburst, while it is reduced during the cool type outburst.

The photometric variability of YY Her was first identified by \cite{1919AN....208..147W}, and further observations by \cite{1932AN....244..289P} and \cite{1939AN....268...71B} contributed to its classification as an irregular variable. 
Later, \cite{1950PASP...62..211H} classified YY Her as a symbiotic star based on its spectral features such as TiO band heads typical of M2-type giants, along with emission lines of H I, He I, He II, and O [III] and noted its close resemblance with other classical symbiotic stars AX Per, CI Cyg, and Z And.
\cite{1991A&A...248..458M} found that the hot component in the system is a WD with a temperature of 10$^{5}$ K and luminosity of 1100 L\textsubscript{\(\odot\)} based on modeling IUE spectra.

A detailed study of the light curves of YY Her from 1890-1996 is discussed in \cite{1997A&A...323..113M}, where it was reported to show four outbursts during 1914-19, 1930-33, 1981-82 and 1993-96, along with six small eruptions during 1890, 1903, 1942, 1954, 1965, and 1974. The authors also estimated the orbital period of the system, P = 590d, with a visual amplitude of \textless{} 0.3 mag. 
A later study by \citet{2006MNRAS.372.1325F} estimated the orbital period of 593.2 days and found a steady decline of the system from 1890 up to 2001, with a rate of 0.01 mag per 1000 days, suggestive of a past symbiotic nova.

The first detailed spectroscopic study of the YY Her outburst (1993) performed by \citet{1997ARep...41..802M}, in which the temperature of the hot component during the outburst was estimated to be $8.5\times 10^{4}$ K, dropping from $1.1\times 10^{5}$ K in quiescence, while the bolometric luminosity of the hot component increased by a factor of $\geq 6$. Further studies by \citet{2000ARep...44..190T} confirm that the hot component of YY Her is accompanied by increased brightness and reduced temperature during the 1993 outburst.
They also suggest that the cool component of YY Her fills its Roche lobe; most of the emission measure of the gaseous envelope is concentrated around the hot component with sharp boundaries, and the system is observed nearly edge-on.
Furthermore, high-resolution observations by \citet{2001AstL...27..703T} during the outburst revealed P Cyg profiles in the He I 5876 \AA{} and 7065 \AA{} lines, with moderate outflow velocities ($\sim 100~\text{km}~\text{s}^{-1}$).

\citet{2001IBVS.5046....1H} published long-term CCD photometric data and identified secondary minima in the light curve, which were later confirmed by \citet{2002AstL...28..620K}. YY Her belongs to a group of symbiotic stars that exhibit secondary minima in their light curves. 
(e.g. CI Cyg, BF Cyg; \citealp{2003ASPC..303..151M}). \cite{2002A&A...392..197M} suggested that secondary minima arise from
a combination of ellipsoidal modulation caused by the distorted giant and sinusoidal variation of the nebular continuum. Alternative explanations to secondary minima, such as WD eclipsing with an optically thick envelope surrounding it \citep{2001IBVS.5046....1H,2006Ap&SS.304..307H} and star spots \citep{2006MNRAS.372.1325F} are also proposed to explain this phenomenon.

Since the 1993 outburst, two additional outbursts were reported in 2003–05 and 2013 \citep{2013ATel.4996....1M}; however, follow-up studies were sparse.
 During February 2021, YY Her underwent a new outburst, reported by \cite{2021ATel14458....1S} in the Astronomer's Telegram. The authors estimated a CV (unfiltered magnitude with V band zero point) of 11.87 mag on 2021 March 15.006 UTC - much brighter than the average quiescent CV of $\sim$ 13 mag. Furthermore, they identified that the rise in magnitude began from 2021 February 25.60 UT from the All-Sky Automated
Survey for Supernovae (ASAS-SN) g-band light curve. Subsequent photometric and spectroscopic observations by \cite{2021ATel14464....1M} revealed a strengthened blue continuum and significantly enhanced emission line fluxes compared to observations taken 583 d earlier, close to the orbital period of YY Her (593.09 d). It was concluded that YY Her is undergoing a hot-type classical symbiotic outburst. Ultraviolet observations from the Swift UVOT instrument by \cite{2021ATel14469....1M} estimated an unabsorbed UV luminosity of 730 L\textsubscript{\(\odot\)} assuming 8.2 kpc distance and interstellar reddening E(B-V) = 0.2 mag. The source remained undetected in X-ray observations taken using Swift Swift X-ray Telescope with an unabsorbed flux upper limit $< 3 \times 10^{-13} \ \text{erg} \ \text{cm}^{-2} \ \text{s}^{-1}$.

We followed up on the hot-type classical symbiotic outburst reported in YY Her using the Himalayan Chandra telescope (HCT) from 2021 to 2023. Long-term follow-up observations covering the outburst and one additional orbital cycle allowed us to understand the evolution of different parameters during the outburst and identify the effect of the orbital phase on their estimation. 

In Section 2, we describe the observational data used in our study. The light curve analysis and orbital period estimation are discussed in Section 3.1. Distance and reddening values are determined in Section 3.2. The spectral energy distribution is discussed in Section 3.3. In Section 3.4, spectral and line evolution is discussed. The temperature and luminosity estimation of the hot component is described in Section 3.5. The discussions and important results are summarized in Sections 4 and 5, respectively.

\section{Observations} \label{sec:obs}

\subsection{Photometry} \label{subsec:phot}

\subsubsection{ASAS-SN, ZTF, and GAIA}

We obtained V- and g-band light curves for YY Her from the ASAS-SN \citep[][]{2014ApJ...788...48S, 2017PASP..129j4502K}, spanning JD 2456371.07 to JD 2460232.63 (2013 March 19 – 2023 October 15), and g-band photometry from the Zwicky Transient Facility \citep[ZTF;][]{2019PASP..131a8003M, 2022ipac.data.I539I} spanning JD 2458204.0 to JD 2460128.93 (2018 March 26 – 2023 July 03). Additionally, we obtained G, G$_{BP}$, and G$_{RP}$ band magnitudes from Gaia DR3 \citep{2016A&A...595A...1G, 2023A&A...674A...1G} covering the period JD 2456920.49 to JD 2457876.99 (2014 September 19 to 2017 May 03). Figure \ref{lc_full} displays the Gaia, ASAS-SN and ZTF light curves. The higher cadence g-band data from ZTF and ASAS-SN overlapped in time.

\subsubsection{GIT} \label{subsec:git}
We carried out photometric observations from the GROWTH-India Telescope\footnote{(\url{https://sites.google.com/view/growthindia/about})} \citep[GIT;][]{2022AJ....164...90K}, a fully robotic 0.7-m telescope located at the Indian Astronomical Observatory (IAO), Hanle, India. Observations are taken in Sloan Digital Sky Survey (SDSS) griz prime 
filters, from JD 2459618.5 to JD 2460116.3 (2022 February 8–2023 June 20), covering the subsequent orbital cycle of YY Her from primary minima after the 2021 outburst. 
The telescope is equipped with a \( 4096 \times 4108 \) Andor iKon-XL CCD, with an image scale of \( 0.67'' \, \text{per pixel} \) and a field of view (FoV) of \( 0.7^\circ \).
The GIT images were pre-processed by an automated pipeline, which performs bias subtraction, flat-field correction, and cosmic-ray removal \citep{2022MNRAS.516.4517K}.
Magnitudes are estimated using PSF photometry and zero-points of individual filters are estimated using Pan-STARRS field observations. The griz-band photometric data are tabulated in Table \ref{git_phot} in the Appendix.

\subsubsection{AAVSO}
We obtained publicly available photometry from the American Association of Variable Star Observers \citep[AAVSO\footnote{\url{https://www.aavso.org}};][]{AAVSODATA} International Database. 
The data consist of CCD and CMOS photometry across various filters, as well as visual estimates. We used V-band photometric data spanning JD 2451760.7 to 2460169.3 (2000 August 4 – 2023 August 12) to construct a long-term light curve for period estimation in our study.

\subsubsection{Photometry from Literature}
\label{lit}

Since the 1993 outburst of YY Her, multiple studies have investigated its photometric behavior and published follow-up photometric data \citep{1997A&A...323..113M,2000ARep...44..190T, 2001AstL...27..703T, 2001IBVS.5046....1H,2006Ap&SS.304..307H, 2002AstL...28..620K, 2002A&A...392..197M}. We compiled CCD and photoelectric based V-band data from these works to construct a long-term light curve. Data from \cite{2006Ap&SS.304..307H} were corrected to account for zero points associated with different observation facilities, as specified in the original paper.

\subsection{Spectroscopy}

Spectroscopic follow-up observations of YY Her outburst were carried out between 2021 March 18 and 2023 August 07. Low-resolution optical spectra were obtained from the Himalayan Faint Object Spectrograph Camera (HFOSC) mounted on the 2-m Himalayan Chandra Telescope (HCT). Spectra cover the wavelength range of 3800 \AA{} to 7500 \AA{} in the grism 7 (Gr7) configuration and 5400 \AA{} to 9000 \AA{} in the grism 8 (Gr8) configuration with a resolution of R $\sim$ 1300 and 2200, respectively. Data reduction was carried out following standard procedures such as bias subtraction, flat fielding, and extraction tasks in the Image Reduction and Analysis Facility (IRAF\footnote{IRAF is distributed by the National Optical Astronomy Observatory, which is operated by the Association of Universities for Research in Astronomy (AURA) under a cooperative agreement with the National Science Foundation.}). We achieve this using the pipeline based on PYRAF modules, which uses the IRAF tasks inside. After wavelength calibration and response correction using standard stars, grism 7 and grism 8 spectra were combined to get the final spectra.  
The combined spectra were converted to an absolute flux scale by matching the synthetic photometry \citep[calculated using the SDSS g prime filter transmission profile from the SVO Filter Profile Service;][]{2012ivoa.rept.1015R} to the ASAS-SN g-band photometry. The calibration adopted the AB magnitude system, with a zero-point flux density of 3631 Jy.
The propagation of errors is carried out in all steps, including the extraction to the absolute flux calibration, and is explained in detail in \cite{2023MNRAS.526.6381S}. The log of observations is given in Table \ref{log}.

\begin{table}
	\caption{Observational log for spectroscopic data obtained for YY Her.}
	\label{log}
	\resizebox{1\hsize}{!}{\begin{tabular}{cccccc}
			\hline

			&  & \textbf{Exposure}  &   \textbf{Wavelength}   \\
			\textbf{Date} & \textbf{JD}  & \textbf{time}  & \textbf{range} \\
			\textbf{yyyy-mm-dd} & & \textbf{(s)} & \textbf{(\AA)} \\
			\hline

                2021-03-18 &    2459292.48 & 180 + 180 & 3600-9000 \\[0.25ex]
                2021-03-20 &    2459294.48 & 120 + 120 & 3600-9000 \\[0.25ex]
                2021-03-22 &    2459296.44 & 180 + 180 & 3600-9000 \\[0.25ex]
                2021-03-24 &    2459298.41 & 120 + 120 & 3600-9000 \\[0.25ex]
                2021-03-26 &    2459300.42 & 120 + 120 & 3600-9000 \\[0.25ex]
                2021-04-03 &    2459308.48 & 120 + 120 & 3600-9000 \\[0.25ex]
                2021-04-07 &    2459312.34 & 120 + 120 & 3600-9000 \\[0.25ex]
                2021-04-25 &    2459330.40 & 120 + 120 & 3600-9000 \\[0.25ex]
                2021-05-01 &    2459336.41 & 180 + 180 & 3600-9000 \\[0.25ex]
                2021-07-17 &    2459413.34 & 180 + 180 & 3600-9000 \\[0.25ex]
                2021-08-17 &    2459444.22 & 360 + 360 & 3600-9000 \\[0.25ex]
                2021-08-24 &    2459451.24 & 180 + 180 & 3600-9000 \\[0.25ex]
                2021-09-01 &    2459459.26 & 180 + 180 & 3600-9000 \\[0.25ex]
                2021-10-13 &    2459501.08 & 180 + 180 & 3600-9000 \\[0.25ex]
                2021-10-22 &    2459510.11 & 180 + 180 & 3600-9000 \\[0.25ex]
                2022-02-11 &    2459621.52 & 180 + 180 & 3600-9000 \\[0.25ex]
                2022-03-08 &    2459647.44 & 300 + 300 & 3600-9000 \\[0.25ex]
                2022-03-21 &    2459660.45 & 300 + 300 & 3600-9000 \\[0.25ex]
                2022-05-09 &    2459709.38 & 180 + 180 & 3600-9000 \\[0.25ex]
                2022-06-11 &    2459742.30 & 180 + 180 & 3600-9000 \\[0.25ex]
                2022-08-30 &    2459822.28 & 240 + 240 & 3600-9000 \\[0.25ex]
                2022-10-01 &    2459854.13 &  240 + 60 & 3600-9000 \\[0.25ex]
                2023-03-02 &    2460006.39 & 900 + 900 & 3600-9000 \\[0.25ex]
                2023-06-10 &    2460106.31 & 180 + 180 & 3600-9000 \\[0.25ex]
                2023-08-07 &    2460164.17 & 180 + 180 & 3600-9000 \\[0.25ex]
			\hline
	\end{tabular}}
\end{table}

\begin{figure*}
\begin{center}
\includegraphics[width=2\columnwidth]{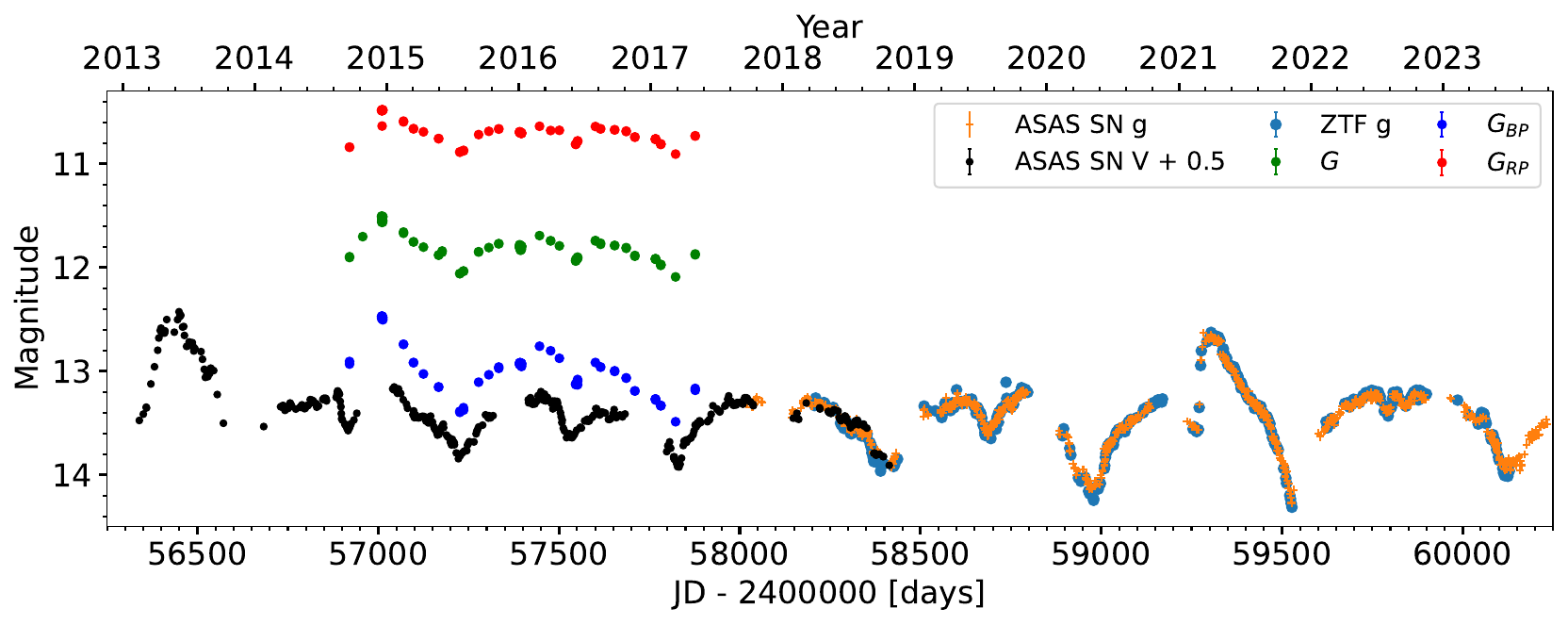}
	\caption{YY Her outburst light curve shown in multiband photometry available from GAIA, ASAS-SN and ZTF. 
    The 2013 and 2021 outbursts are visible in the light curve. An offset is applied to ASAS-SN V band for clarity.}
	\label{lc_full}
\end{center}
\end{figure*}

\section{RESULTS}

\subsection{Optical light curve of YY Her}

\subsubsection{Periodic Behavior}

The optical light curve of YY Her from 2013 to late 2023, consisting of data from Gaia, ASAS-SN, and ZTF, is shown in Figure \ref{lc_full}. A 0.5 mag offset is applied to the ASAS-SN V band. Primary and secondary minima are visible in the light curve.  
We estimated the periodicity of the light curve using Lomb-Scargle periodogram \citep[LSP;][]{1976Ap&SS..39..447L, 1982ApJ...263..835S} and obtained periods of 579.19 d and 578.92 d corresponding to the highest peaks in the ASAS-SN g and ZTF g bands, and a period of 525.77 d in a closer peak in the ASAS-SN V band. Similarly, we obtained periods of 604.71 d for the AAVSO V band and 583.61 d corresponding to the nearest peak for previously published V-band measurements (hereafter Literature V; see Section \ref{lit}).

\begin{figure}
\centering
\includegraphics[width=0.5\textwidth]{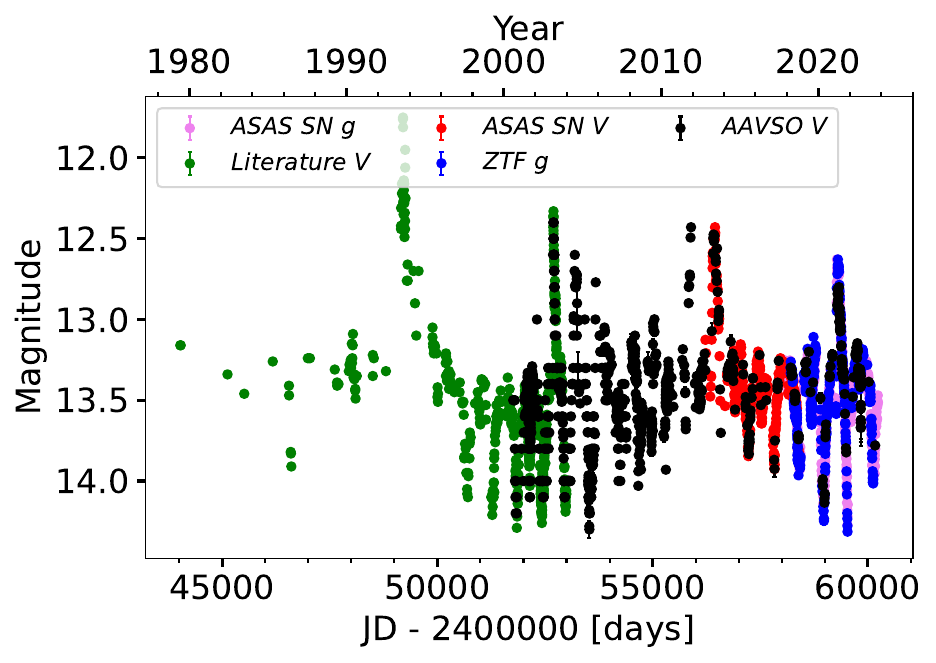}
	\caption{Combined V and g band data from ASAS-SN, ZTF, AAVSO and literature after applying appropriate shift (see text for details).}
	\label{plot:comb}
\end{figure}

\begin{figure}
\begin{center}
\includegraphics[width=\columnwidth]{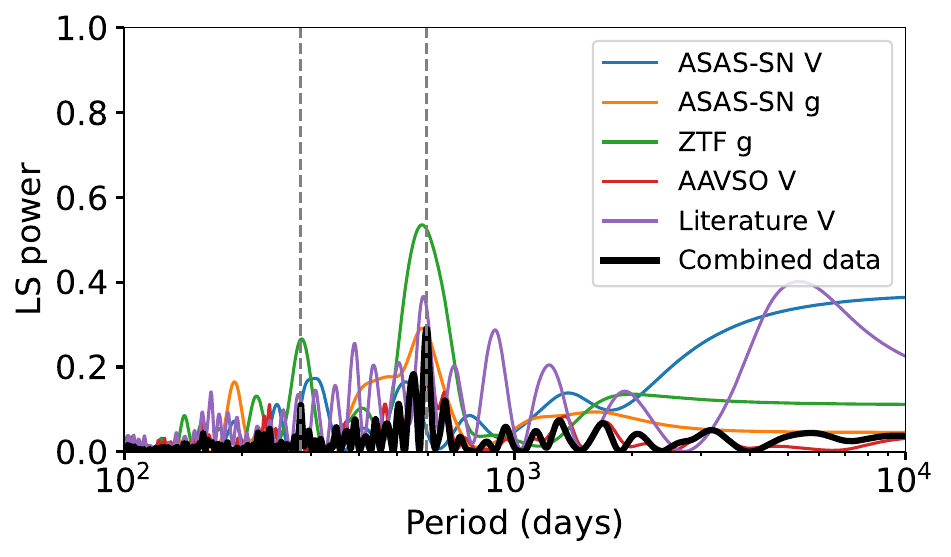}
	\caption{LSP of YY Her using ASAS-SN g and V band, ZTF g band, AAVSO V band, V data compiled from the literature  and combined data is plotted (see text for details). The period corresponding to primary minima and secondary minima is marked with dashed lines. }
	\label{lsp}
\end{center}
\end{figure}

To improve period estimation, we constructed a long-term light curve combining photometric data from ASAS-SN g, ZTF g, ASAS-SN V, AAVSO V, and Literature V measurements (see Figure~\ref{plot:comb}). A 0.5 mag shift was applied to all V band data to align it with the g
band. The LSP analysis of the combined data set shows a major peak at 594.08 d, corresponding to the primary minima, and a secondary peak at 282.95 d, associated with the secondary minima. Both peaks have a false alarm probability of $<0.01$ percent. 
The resulting periodograms are presented in Figure \ref{lsp}.

Using the period of primary minima obtained from LSP, we initially fitted a single sinusoidal model and derived a period of 594.27 ± 0.5 d.
We also attempted to fit a combination of two sinusoidal curves (hereafter double sinusoidal model) representing periods of primary minima and secondary minima (see Figure \ref{plot:double}).
The double sinusoidal model provides a lower chi-square fit compared to a single sinusoidal model and we estimated a period of 593.09 $\pm$ 0.30 days, which is in agreement with the previous estimate of \cite{2006MNRAS.372.1325F} within the margin of error and closer to the period estimation based on the long-term light-curve study by \cite{1997A&A...323..113M}. Hence, we adopted the ephemeris based on double-sinusoidal fit given by Equation \ref{eq:ephemeris} to calculate the orbital phase for the rest of the paper.

\begin{equation}
\label{eq:ephemeris}
JD_{min} = 2458968.13 \pm 2.04 + 593.09 \pm 0.30 \times E
\end{equation}

Phase (\( \phi \)) = 0.0 corresponds to the inferior conjunction of the cool giant, coinciding with the primary minima, while \( \phi = 0.5 \) corresponds to the superior conjunction. Hereafter, when the phase is written within the range $ 0 \leq \phi \leq 1$, it corresponds to a repeating phase interval unless specified with an exact date (e.g., 0.5 represents 0.5, 1.5, 2.5, etc.).

Gaia G$_{BP}$, G and G$_{RP}$ light curve indicates that during the minimum, the largest amplitude is observed in the G$_{BP}$ band (blue), while it is weaker in the G$_{RP}$ band (red). This trend aligns with observations in other shell-burning symbiotic stars, where irradiation from burning WD significantly influences the blue spectral region \citep[e.g. TCP J18224935-2408280;][]{2023MNRAS.526.6381S}. 
The modulation can be explained by a combination of the reflection effect  \citep{1970Ap......6...22B, 1986syst.book.....K} and orbital variation in nebular contribution \citep{2001A&A...366..157S, 2002A&A...392..197M}.
%both of which result from irradiation by the hot WD.}

In Figure \ref{lc_full}, we observe secondary minima in the light curve throughout all orbital cycles from 2014. In addition, we note that the secondary minima appear slightly shifted toward \( \phi \sim \)  0.45 (see Figure \ref{lc_git}).  The secondary minima seen in YY Her are explained by the ellipsoidal effect of the cool giant \citep{2002A&A...392..197M}. Other alternative explanations include eclipsing by the hotter component surrounded by an optically thick envelope that mimics a main-sequence star in terms of radius \citep{2006Ap&SS.304..307H}. 
We favor the interpretation of the ellipsoidal effect as the primary cause of the observed phenomena. The double sinusoidal model we used to fit the combined light curve is similar to the phenomenological model proposed by \cite{2002A&A...392..197M}. However, we noticed that the secondary minima are steeper than in our model. We further explore the cause of secondary minima in Section \ref{Ca_II}, where we analyze the variation of the Ca II line during the orbital cycle.

\subsubsection{Outburst Light Curve}

YY Her underwent two outburst events in the last ten years -- 2013 \& 2021 \citep{2013ATel.4996....1M,2021ATel14458....1S}.  ASAS-SN and ZTF g band, show that the YY Her 2021 outburst began around 2021 February 25 and peaked at 12.62 mag with an approximately 1 mag brightening event. The outburst coincided with secondary minima, which hindered an accurate estimation of the amplitude of the outburst. 2013 outburst in the V-band light curve also shows a brightening of $\sim$ 1 mag. Compared to the 1994-96 outburst reported in YY Her \citep{1997ARep...41..802M}, the 2021 outburst had a lower amplitude. Hot-type outbursts typically show a lower optical brightening than cool-type classical symbiotic outbursts. This is also seen in previously reported AG Dra outbursts \citep{AG_DRA_Gonzalez_1999}.

Interestingly, the primary minima observed before and after the 2021 outburst showed a more significant reduction in brightness (dimmer by $\sim$ 0.2 mag) compared to other primary minima in the light curve. Similar behavior was observed by \cite{1997A&A...323..113M} during previous outbursts. This enhanced depth of minima may stem from increased stellar wind activity originating from the giant star during this period.

\begin{figure}
\begin{center}
\includegraphics[width=\columnwidth]{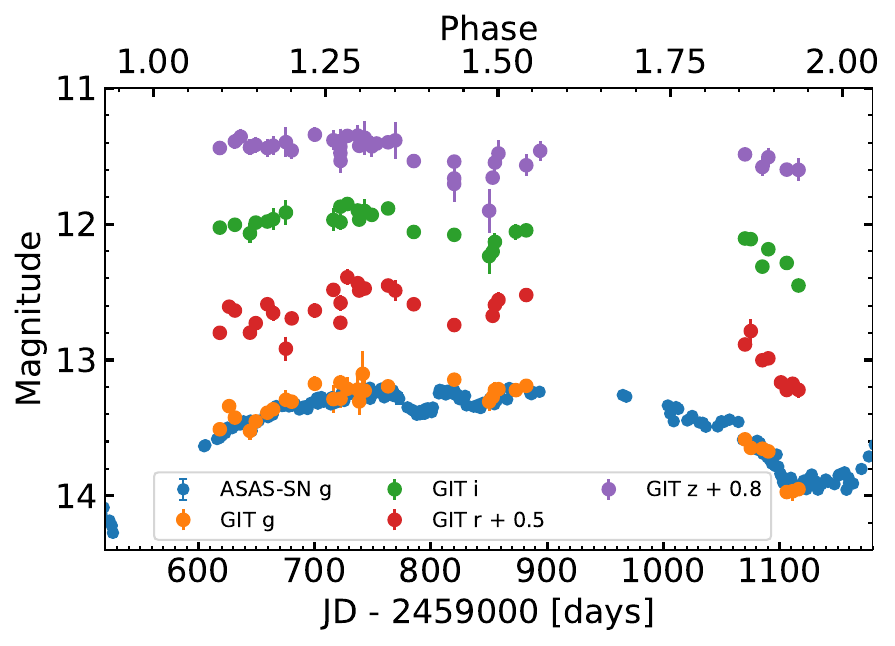}
	\caption{YY Her light curve shown in GIT multi-band photometry. Phase estimated based on the ephemeris given in Equation \ref{eq:ephemeris}. Offsets are applied for clarity.}
	\label{lc_git}
\end{center}
\end{figure}

From the ASAS-SN and ZTF g-band light curves, we observed a notable reduction in the secondary minima in the subsequent orbital cycle following the outburst. In fact, it looked like a smaller peak inside a secondary minima (see Figure \ref{lc_full} $\&$ \ref{lc_git}). The GIT light curve obtained during the same period suggests that the increase in magnitude during secondary minima is only significant in the g band, not in the r, i, z bands (see Figure \ref{lc_git}). After the 2021 outburst, the hot component (WD) remains significantly heated, causing an increase in magnitude in the blue band (g) during \( \phi \sim \) 1.45, when the hot component is visible to the observer. This should be taken with caution, as we have hardly any photometric points during secondary minima in GIT photometry. 
A similar brightening in the g band light curve following the classical symbiotic outburst in TCP J18224935-240828 is reported by \citep{2023MNRAS.526.6381S}; however, for YY Her, it is limited to the secondary minima, possibly due to the high inclination.

\subsection{Distance and Reddening}
\label{sec_3.3}

In our study, we used Gaia EDR3 parallaxes to determine the geometric and photo-geometric distances to YY Her, using the method by \cite{2021AJ....161..147B}. The geometric distance was calculated as $6.9${\raisebox{0.5ex}{\tiny$\substack{+1.3 \\ -1.2}$}} kpc, while the photo-geometric distance was found to be $5.4${\raisebox{0.5ex}{\tiny$\substack{+0.8 \\ -0.6}$}} kpc. Our SED fitting, discussed in Section \ref{sed}, provided a distance estimate of $5.2${\raisebox{0.5ex}{\tiny$\substack{+0.4 \\ -0.3}$}} kpc, close to the photo-geometric distance.  
The goodness of fit of the astrometric model is -1.5 for Gaia EDR3. A lower than 3 is considered a good fit\footnote{(\url{https://gea.esac.esa.int/archive/documentation/GDR2/Gaia_archive/chap_datamodel/sec_dm_main_tables/ssec_dm_gaia_source.html})}. Gaia distance estimation is based on a Bayesian probabilistic approach which uses prior constructed from a three-dimensional model of our Galaxy assuming that all sources are single stars.
This may introduce significant uncertainties when estimating distances for binary systems. 
We have adopted the photo-geometric distance given \cite{2021AJ....161..147B} in this paper.

Reddening map by \cite{2011ApJ...737..103S} gives a visual extinction value, $A_{\text{v}} = 0.26${\raisebox{0.5ex}{\tiny$\substack{+0.2 \\ -0.2}$}}  in the direction of YY Her. In addition, we also determined visual extinction $A_{\text{v}} = 0.29${\raisebox{0.5ex}{\tiny$\substack{+0.1 \\ -0.1}$}} in direction of YY Her for the estimated Gaia photo-geometric distance, using the 3D map of interstellar dust reddening by \cite{2019ApJ...887...93G}. The value of $A_{\text{v}}$ did not show significant differences beyond 1.3 kpc. The error associated with $A_{\text{v}}$ is determined from the posterior distribution. In our paper, we adopt the value  $A_{\text{v}} = 0.29${\raisebox{0.5ex}{\tiny$\substack{+0.1 \\ -0.1}$}}. Reddening corrections are applied using the extinction law proposed by \cite{1999PASP..111...63F}.

\subsection{Spectral Energy Distribution of Cool Component in YY Her}
\label{sed}

\begin{figure}
\includegraphics[width=\columnwidth]{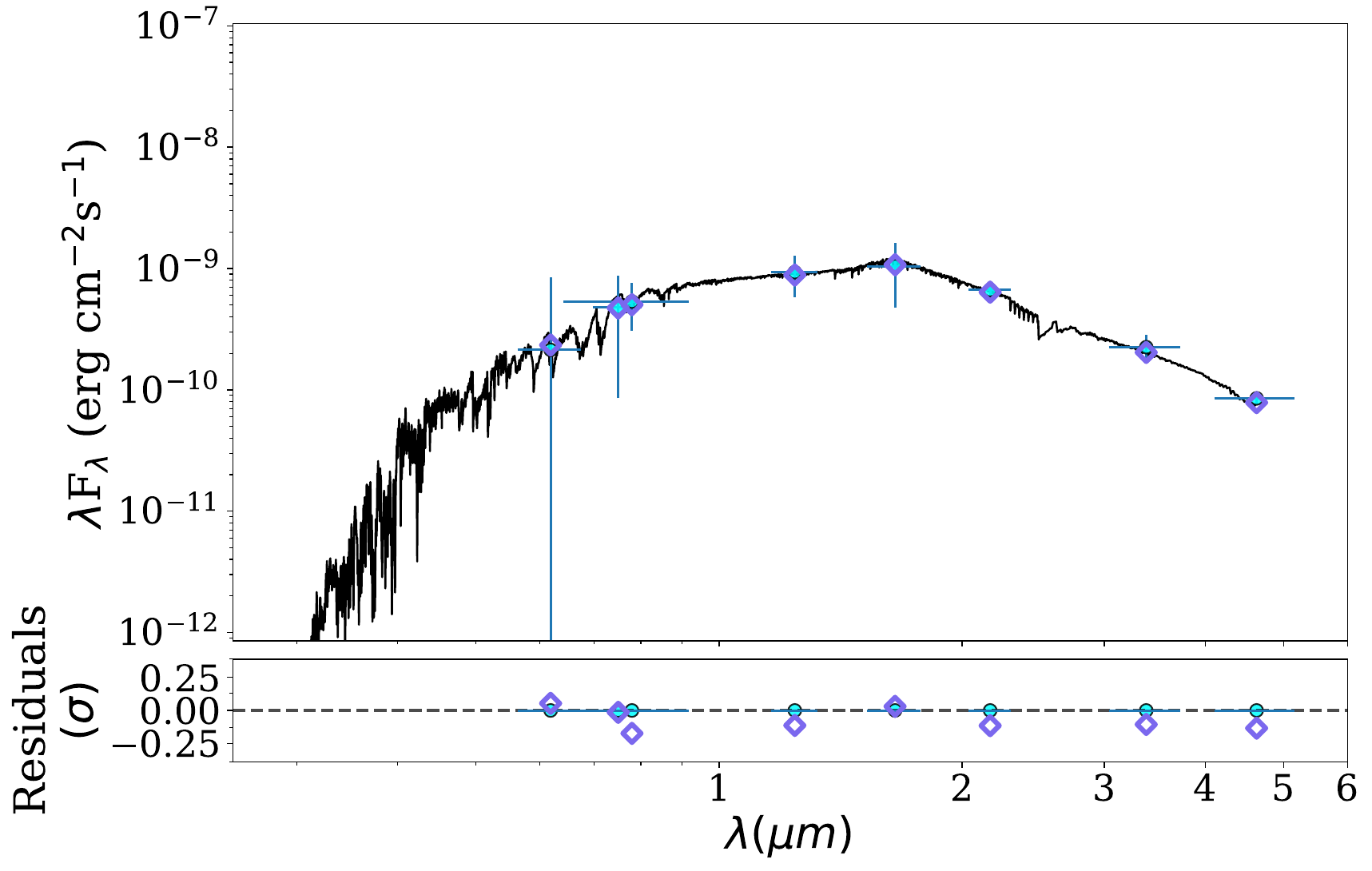}
	\caption{Spectral energy distribution of the YY Her obtained from different photometric bands. (see Section \ref{sed} for details)}
    \label{plot:sed}
\end{figure}

\begin{figure}
\includegraphics[width=\columnwidth]{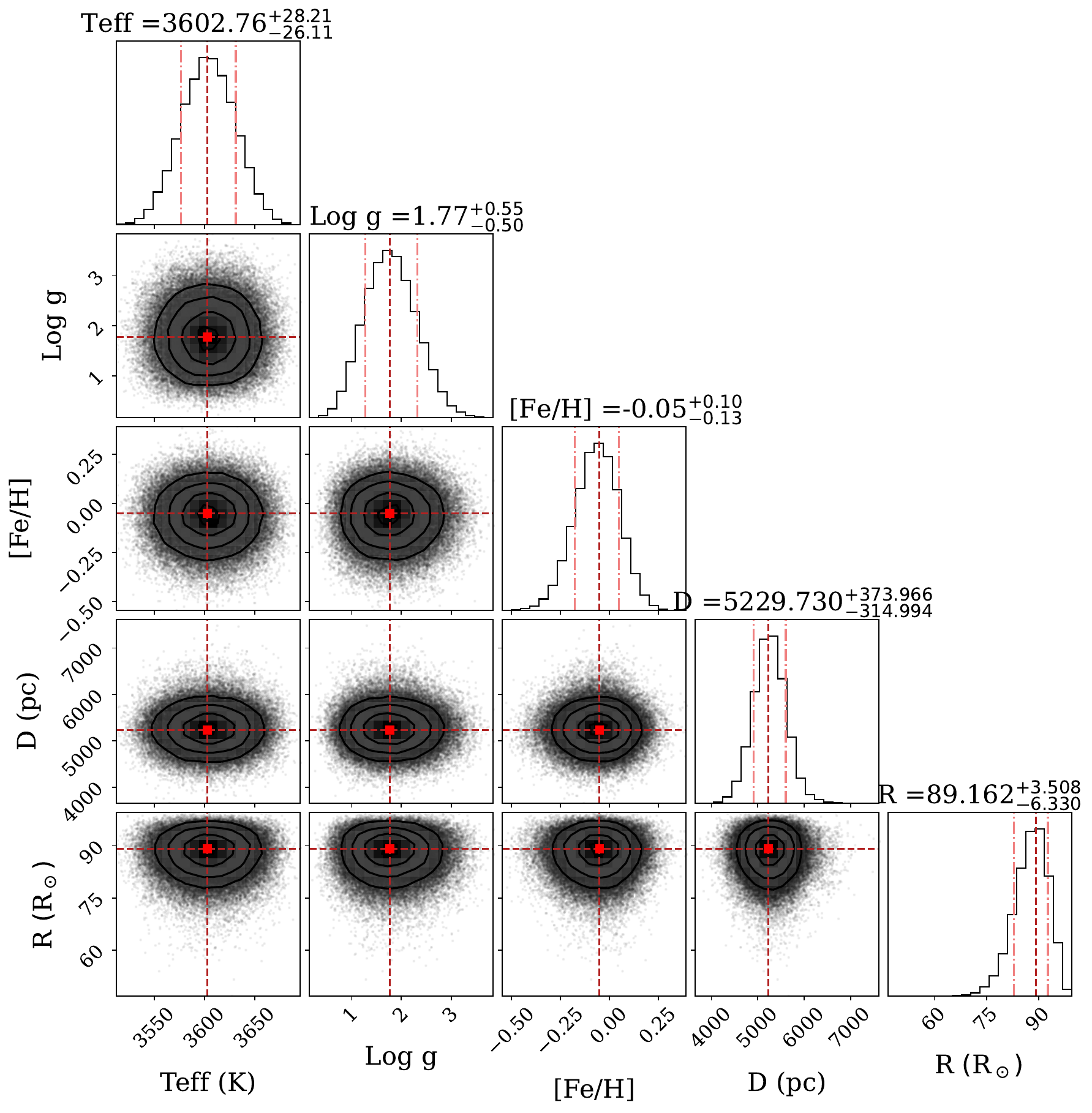}
	\caption{Corner plot showing various parameters derived from SED fitting using synthetic model atmospheres.}
	\label{corner_plot}
\end{figure}

\begin{figure*}
\centering
\includegraphics[width=2\columnwidth]{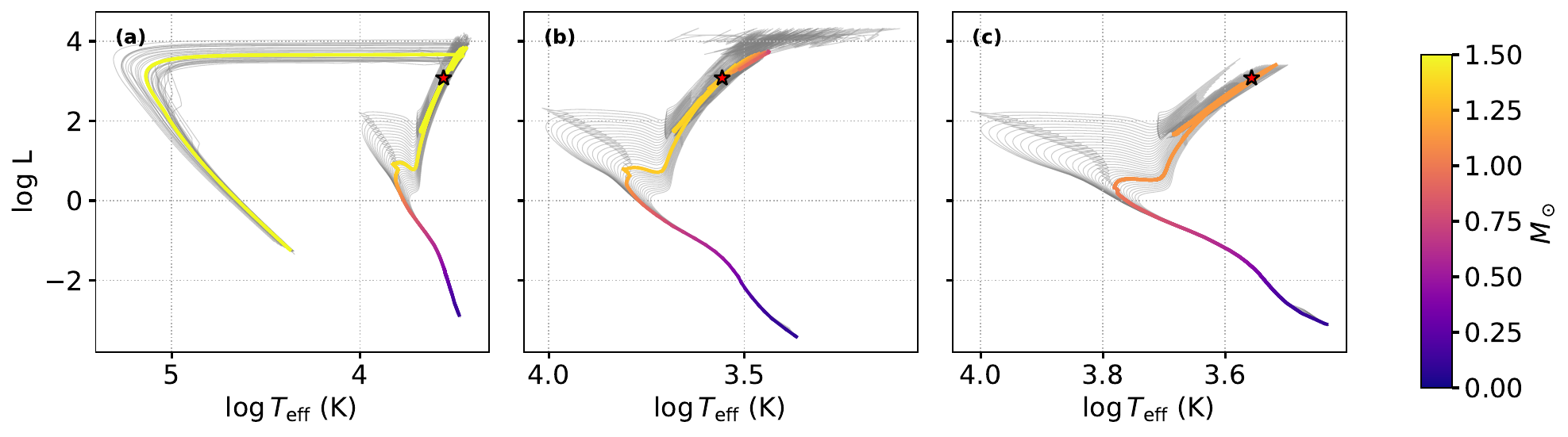}
	\caption{H-R diagrams showing stellar isochrones from (a) MIST, (b) PARSEC, and (c) BaSTI models. The red star with error bars marks the observed effective temperature and luminosity of the cool giant derived from the SED fit.}
	\label{plot:hr_new}
\end{figure*}

\cite{1950PASP...62..211H} initially assigned a spectral type of M2 to the cool giant in YY Her. Subsequent studies by \cite{1987AJ.....93..938K} classified the secondary as a M3-type cool giant, while \cite{1997ARep...41..802M} estimated it to be of M4 type.

We determined the spectral type by constructing an SED using photometric data from Gaia DR3 (G$_{RP}$; \citealp{2016A&A...595A...1G,2023A&A...674A...1G}), APASS ($r'$, $i'$; \citealp{2014CoSka..43..518H}), Two Micron All Sky Survey ($J$, $H$, $K_s$; \citealp{2006AJ....131.1163S}), and Wide-field Infrared Survey Explorer (W1, W2; \citealp{2010AJ....140.1868W}) (see Figure~\ref{plot:sed}).
The cool component in the symbiotic star dominates in redder wavelengths, whereas the contribution from the nebula and WD is higher in the bluer wavelengths. Hence, we only considered filters above 6000 \AA{} in our SED. The SED modelling is performed using the ARIADNE\footnote{(\url{https://github.com/jvines/astroARIADNE})} \citep{2022MNRAS.513.2719V}, which uses synthetic model atmospheres, including PHOENIX v2 \citep{PHOENIX_v2}, BT-NextGen \citep{NextGen_1999, Allard_et_al_2012}, BT-Settl \citep{Allard_et_al_2012}, and BT-Cond \citep{Allard_et_al_2012}, employing Bayesian model averaging techniques to optimally constrain parameters.

We used temperature prior based on the Gaia temperature estimate and its upper limit. The distance prior is taken from geometric distance estimate from  Gaia
(see section \ref{sec_3.3}), with the highest error serving as the upper limit. The $A_{\text{v}}$ value is fixed at 0.29. Additionally, we employ a uniform prior for log g ranging from 0 to 4, derived from our initial fitting, resulting in a radius estimation within the giant star regime. Default priors are used for radius and metallicity. The best fit yields a temperature of $3602.76${\raisebox{0.5ex}{\tiny$\substack{+28.21 \\ -26.11}$}} K, log g = $1.77${\raisebox{0.5ex}{\tiny$\substack{+0.55 \\ -0.50}$}}, radius = $89.16${\raisebox{0.5ex}{\tiny$\substack{+3.51 \\ -6.33}$}}  R\textsubscript{\(\odot\)} and luminosity = $1248${\raisebox{0.5ex}{\tiny$\substack{+287.6 \\ - 215.7}$}} L\textsubscript{\(\odot\)}.
These results correspond to a cool giant of M2 spectral type. 
The corresponding SED parameters are shown in Figure \ref{corner_plot}. 
The mass of the cool giant is estimated by interpolating the best-fit temperature and luminosity from the SED analysis within stellar isochrones (see Figure \ref{plot:hr_new}). The resulting masses are 1.50 $M_\odot$ from MIST \citep{mist_1}, 1.29 $M_\odot$ from PARSEC \citep{2012MNRAS.427..127B}, and 1.14 $M_\odot$ from BaSTI \citep{2018ApJ...856..125H}.

Considering the phase-dependent variability observed in the YY Her spectra during different observations (see Section \ref{optical_spec}), the SED obtained only explains global mean of the photometric observations taken. 
Since we are not modeling and removing the nebular contribution to the SED, the spectral type we have obtained serves as the upper limit. In Addition, the spectral type is likely overestimated because the hot WD irradiates the surface of the cool giant. Furthermore, it should be noted that the SED can more accurately constrain parameters such as temperature, radius, and log g, while [Fe/H] serves as a qualitative measure.

\subsection{Evolution of Optical Spectra}
\label{optical_spec}

\begin{figure*}
\begin{center}
\label{spec2}
    \includegraphics[width=2\columnwidth]{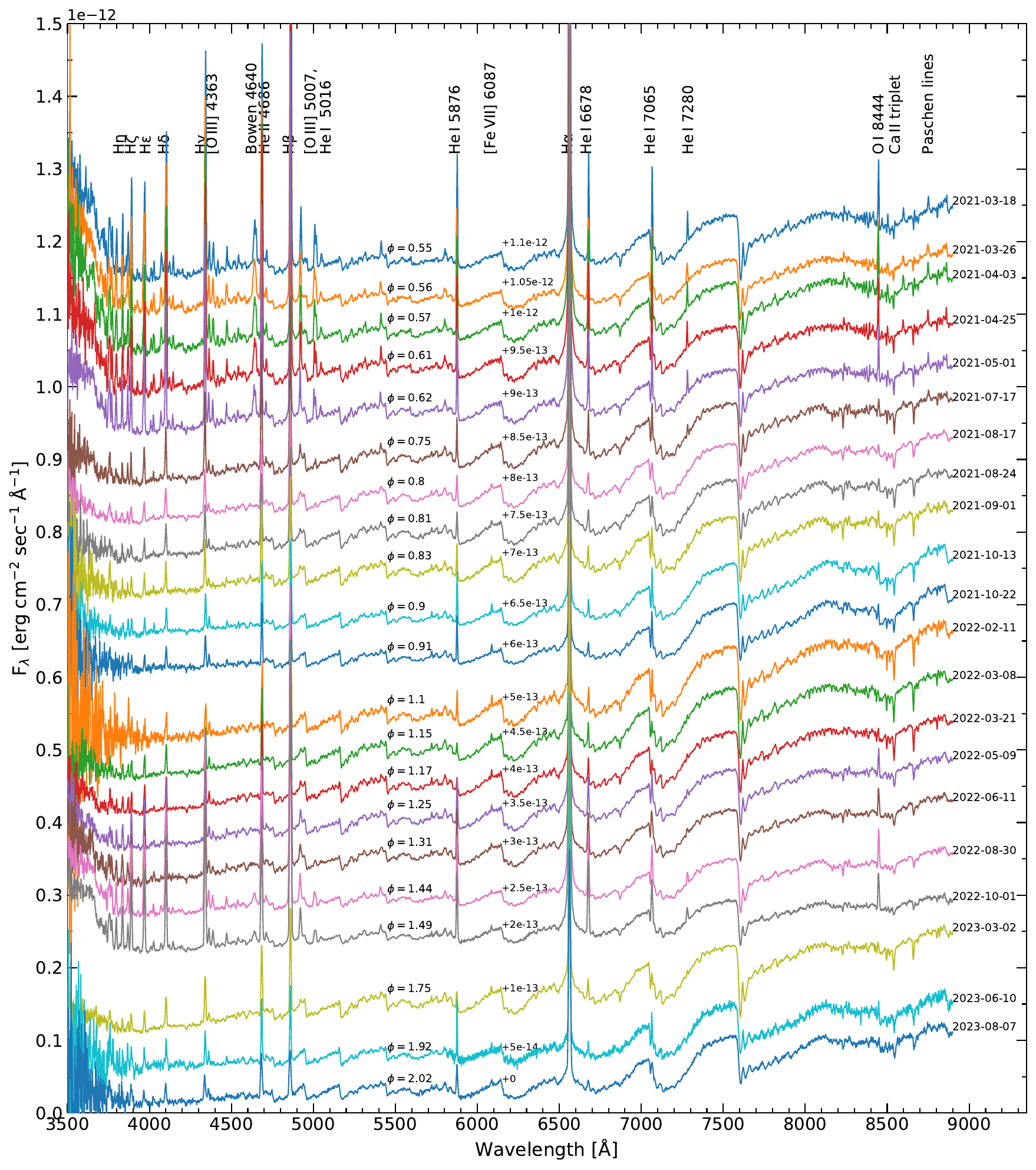}
	\caption{Optical low-resolution spectral evolution of YY Her obtained from 2021 to 2023 covering the outburst and subsequent orbital cycle of YY Her. The identified lines are labeled on the top.  Observation date and orbital phase ($\phi$) are given for each spectrum. Spectra are shifted to the indicated amount for better visibility. Few observed spectra taken in closer epochs are omitted from this figure; they are plotted in the Appendix (see Figure \ref{spectra2}). }
	\label{spectra1}
\end{center}
\end{figure*}

Spectroscopic monitoring of YY Her covered the declining phase of the 2021 outburst and one subsequent orbital cycle. The outburst coincides with the secondary minima in the light curve, and observations were taken from \( \phi = 0.55 \) (18 March 2021) to \( \phi = 2.02 \) (07 August 2023). The follow-up spectroscopic observations are plotted in Figure \ref{spectra1}; a few additional spectra taken at closely spaced epochs--providing no new phase coverage--are included in Appendix \ref{spectra2}.
The optical spectra of YY Her show Balmer series lines, He I, O I, high-excitation lines such as He II, and TiO band heads from the cool component throughout the evolution, including outburst, quiescence, and both primary and secondary minima. We also noted that the Raman-scattered O VI band is absent in the spectrum, and O[III], Bowen 4640 lines are present only in particular phases.

\begin{figure}
\includegraphics[width=\columnwidth]{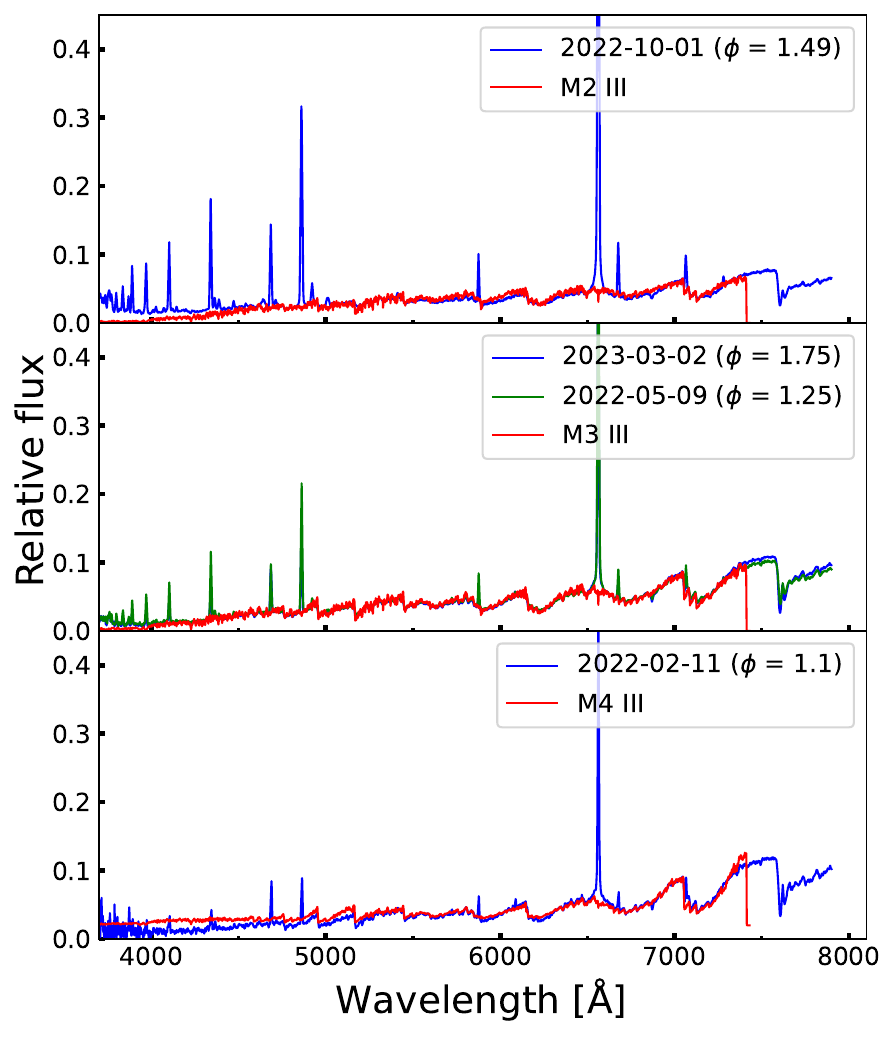}
	\caption{Spectral type comparison plot with phase 0,  0.25/0.75 and 0.5 with respect to M2 III, M3 III, and M4 III spectral type from MILES library. Spectra is scaled for comparison.}
	\label{spec:type}
\end{figure}

Optical spectra during different orbital phases show significant changes in spectral profile (see Figure \ref{spectra1}). We further studied this by comparing it with reference spectra of the different spectral types obtained from the MILES library \citep{2011A&A...532A..95F}.
We found that the spectrum obtained at \( \phi =\) 1.49 (2022 October 01) was closely similar to the M2 III, while at \( \phi =\) 1.25 and 1.75 (2022 May 09 and 2023 March 02) resembled M3 III (see Figure \ref{spec:type}). At \( \phi =\) 1.1 (2022 February 11) the spectrum is closely aligned to M4 III. The surface of the cool giant facing the WD has a higher temperature, which is reflected in the estimated spectral types in different phases. During the inferior conjunction of the cool giant, the heated region is least visible and shows a later spectral type (M4 III), while during the superior conjunction, it shows an earlier spectral type (M2 III).

\begin{figure*}
\centering
\includegraphics[width=2\columnwidth]{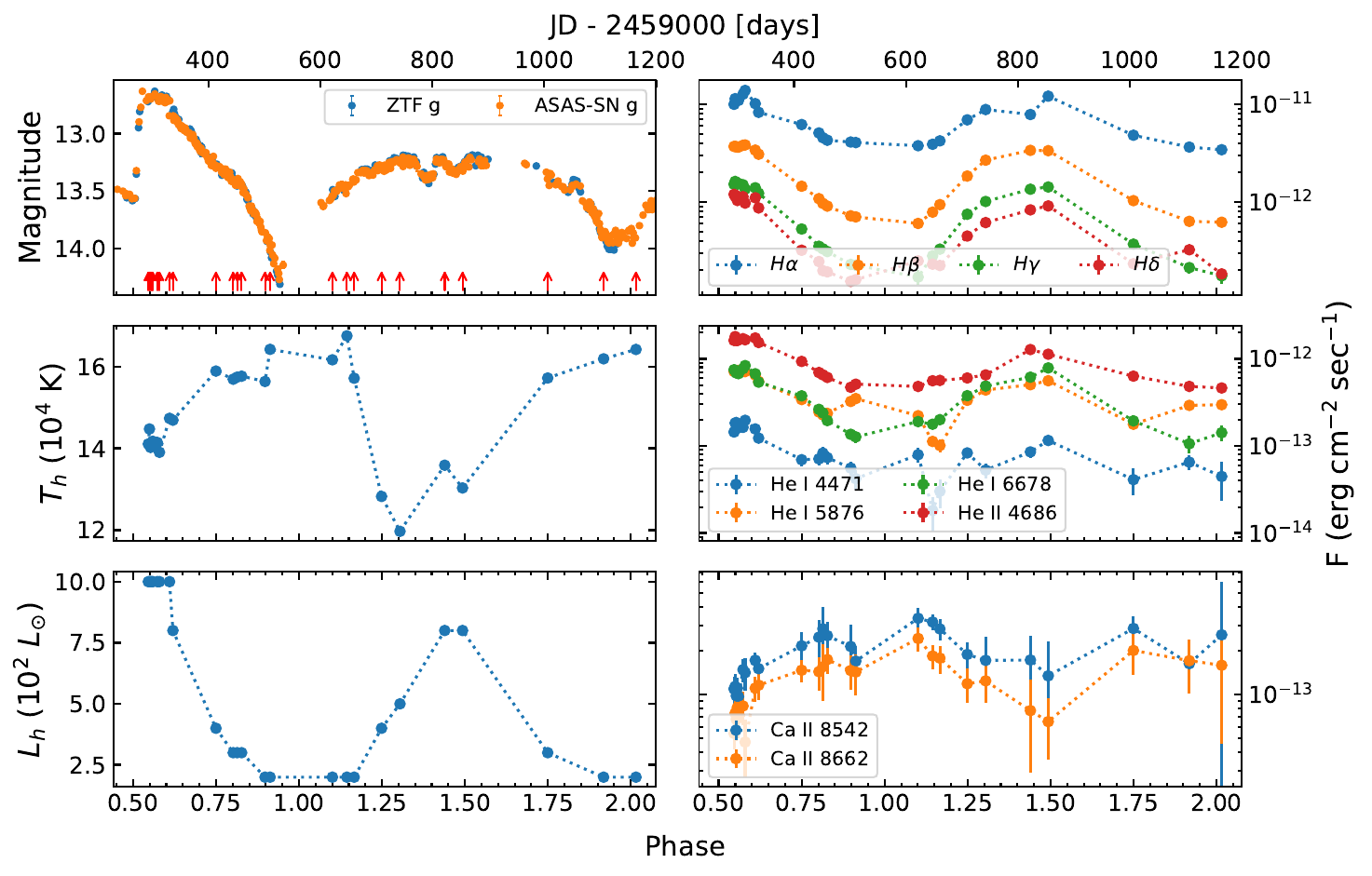}
	\caption{(Top-left) ASAS-SN and ZTF g-band light curve with red arrows indicating the epochs of spectroscopic observations. The subsequent plots show the evolution of the temperature and luminosity of the hot component. The second column shows the evolution of line fluxes. }
 	\label{line:param}
\end{figure*}

\begin{figure}
\includegraphics[width=\columnwidth]{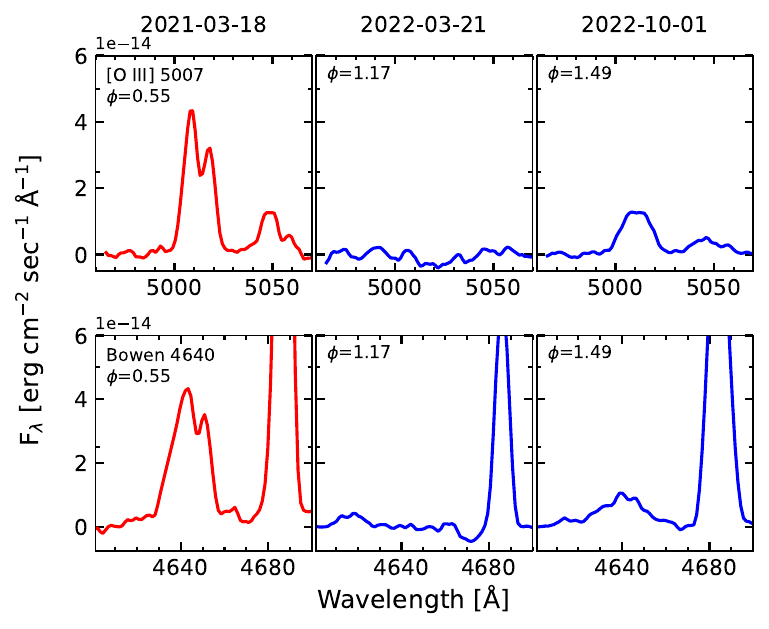}
	\caption{Evolution of O[III] 5007 and Bowen 4640 lines in different phases are plotted (left to right) after subtracting the local continuum. The red color represents the outburst, whereas blue plots are during the next orbital cycle of YY Her. Interestingly, these lines are only visible at  \( \phi = \)  0.55 and 1.49, corresponding to secondary minima, and they strengthened considerably during the outburst.}
	\label{line_o3}
\end{figure}

We measured fluxes of all important lines; the de-reddened values are listed in Table \ref{flux_table}. The flux evolution of the Balmer and He emission lines, and Ca II absorption lines is shown in Figure~\ref{line:param}. In the first spectrum, we obtained during the outburst on \( \phi = 0.55 \) (18 March 2021), emission lines of He II, He I, and O I are significantly strengthened compared to the observation taken on 2022 October 01 at similar \( \phi = 1.49 \), following one orbital cycle. The Bowen 4640 \AA{} and [O III] 5007~\AA{} lines are also strengthened (see Figure \ref{line_o3}). 
The outburst resulted in an enhanced blue continuum and a significantly strengthened Balmer jump, indicating a rise in the temperature of the ionizing source.
We observed an increase in the He II / H$\beta$ ratio during the outburst (\( \phi = 0.55 \)) compared to the similar phase (\( \phi = 1.49 \)) after one orbital cycle. \cite{2021ATel14464....1M} reported similar observations comparing outburst and pre-outburst spectra taken almost 583 days earlier.  
An increase in the He II/H$\beta$ ratio indicates that the outburst is of hot type.

Spectroscopic follow-up observations show the sinusoidal variability of the Balmer lines (Figure \ref{line:param}), which peak around \( \phi = 0.5 \). He I 6678 \AA{} and He II 4686 also show a similar trend, but the amplitude variation is lower than the Balmer lines. He I 7280 \AA {} disappears completely near primary minima, a pattern also evident in O I line 8444 \AA{}.
A prior study by \cite{1989A&A...208...63M} on multiple S-type symbiotic stars using phase-resolved IUE spectroscopy found that UV emission lines strengthened significantly near \( \phi = 0.5 \). \cite{1996ApJ...471..930P,1998ApJ...501..339P} explained this phase-dependent variation in emission line strengths using non-LTE illumination models of the red-giant wind.

\begin{figure}
\centering
\includegraphics[width=0.5\textwidth]{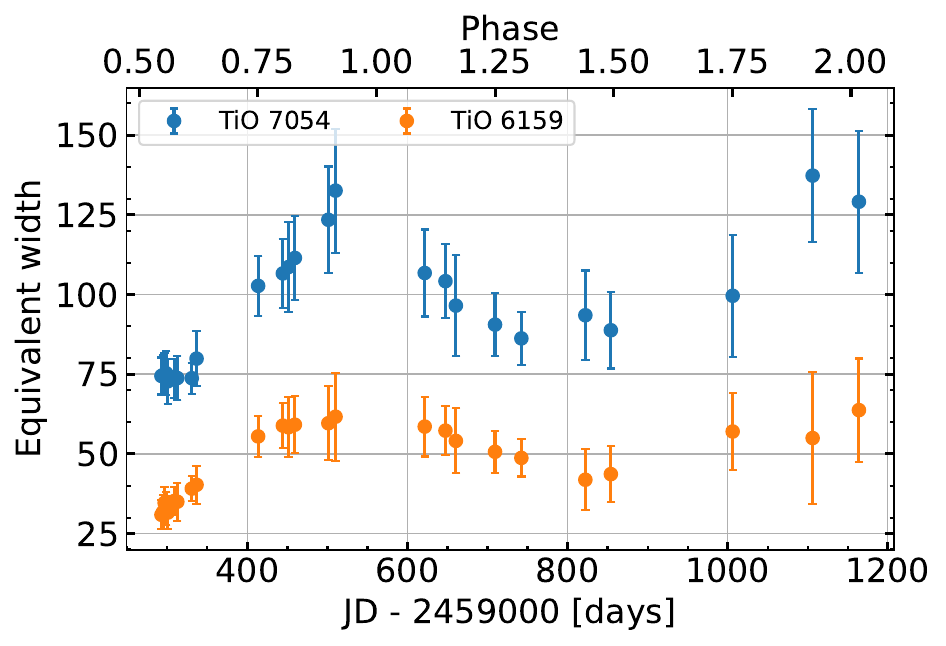}
	\caption{ Equivalent widths of TiO bands at 7054\AA{} and 6159\AA{} plotted against time. The upper x-axis shows the orbital phase.}
	\label{plot:tio}
\end{figure}

The TiO bands appear weaker around \( \phi = 0.5 \) and strengthen as they approach \( \phi = 0 \). 
We estimated the equivalent widths of TiO bands at 7054\AA{} and 6159\AA{} after removing the emission lines and plotted them against the phase (see Figure \ref{plot:tio}). TiO bands show a periodic variation that arise from the inverse relationship between the temperature and band strength, which is consistent with the spectral types observed at different phases. 

The O [III] 5007 \AA{} is clearly identifiable during secondary minima (\( \phi = 0.5 \)) (see Figure \ref{line_o3}). The O [III] line was blended with a nearby He I 5016 \AA{} line, and during quiescence (\( \phi = 1.49 \)), the line strength was reduced. Although it remains difficult to observe, the strength of the O [III] line is progressively reduced while moving away from the secondary minima. Notably, the Bowen 4640 \AA{} line is visible only in a very narrow phase interval (\( \phi = 0.5 \pm 0.1\)) close to the secondary minima. 
We identified the [Fe VII] 6083 \AA{} line in all phases except during the outburst. This behavior is consistent with observations from outbursts in other symbiotic stars (e.g., TCP J18224935-2408280 -- \citealp{2023MNRAS.526.6381S}; AX Per -- \citealp{2011A&A...536A..27S}; FN Sgr -- \citealp{2005A&A...440..239B}).

\subsubsection{Ca II line variation}
\label{Ca_II}
Ca II triplet absorption lines originate from the photosphere of the cool giant in YY Her. We estimated the flux of relatively stronger and less contaminated Ca II 8542 \AA{}, Ca II 8662 \AA{} lines. Ca II lines show variability with half of the orbital period of the system, whereas the Balmer lines and the He lines follow the orbital periodicity (see Figure \ref{line:param}).  The origin of such a period is due to the ellipsoidal modulation of the giant star. In Figure \ref{line:param}, we can see that the Ca II lines are strengthened near \( \phi = \) 0.25 and 0.75, while they weakened closer to \( \phi = \) 0.0 and 0.5.

\subsection{Temperature and luminosity of the hot component}

\begin{table*}
	\centering
	\caption{ The de-reddened absolute fluxes of H$\beta$, He II \,4686\,\AA{}, He I 4471\,\AA{} and He I 5876\ \AA{} together with the estimated luminosity, temperature
and radius of the YY Her hot component.}
	\label{tab:hot}
	\begin{tabular}{ccccccccccc} 
		\hline		
\multicolumn{1}{c}{Date}&\multicolumn{1}{c}{JD}& \multicolumn{1}{c}{Phase}&\multicolumn{
4}{c}{Flux \,ergs\,cm$^\mathrm{-2}$\,s$^\mathrm{-1}$}& 
\multicolumn{1}{c}{$T_\mathrm{ h}$}&
\multicolumn{1}{c}{$L_\mathrm{h}$}&
\multicolumn{1}{c}{$R_\mathrm{h}$}\\ 

yyyy-mm-dd&\multicolumn{1}{c}{since 2400000}& &\multicolumn{1}{c}{He II\,4686\,\AA} & 
\multicolumn{1}{c}{H$\beta$} & \multicolumn{1}{c}{He I 4471\,\AA} & \multicolumn{1}{c}{He I 5876\,\AA}  &\multicolumn{1}{c}{[10$^3$\,K]} & \multicolumn{1}{c}{[$L_{\sun}$]} & 
\multicolumn{1}{c}{[$R_{\sun}$]} \\
		\hline

2021-03-18  & 59292.48& 0.55& 1.63e-12  &   3.7e-12 & 1.45e-13 &   7.18e-13     & 141.1&    1020.0& 0.054\\ 
2021-03-20  & 59294.48& 0.55& 1.8e-12  &   3.7e-12 & 1.83e-13 &   7.24e-13     & 144.7&    1070.0& 0.052\\ 
2021-03-22  & 59296.44& 0.55& 1.63e-12  &  3.68e-12 & 1.86e-13 &    7.3e-13     & 140.2&    1030.0& 0.054\\ 
2021-03-24  & 59298.41& 0.56& 1.62e-12  &  3.65e-12 &  1.6e-13 &   7.34e-13     & 141.1&    1020.0& 0.053\\ 
2021-03-26  & 59300.42& 0.56& 1.65e-12  &  3.65e-12 & 1.72e-13 &   6.93e-13     & 141.7&    1020.0& 0.053\\ 
2021-04-03  & 59308.48& 0.57& 1.7e-12  &   3.8e-12 & 1.64e-13 &   7.18e-13     & 141.4&    1060.0& 0.054\\ 
2021-04-07  & 59312.34& 0.58& 1.65e-12  &  3.84e-12 & 1.97e-13 &   7.13e-13     & 139.0&    1060.0& 0.056\\ 
2021-04-25  & 59330.4& 0.61& 1.74e-12  &   3.4e-12 & 1.57e-13 &   6.53e-13     & 147.3&    1000.0& 0.049\\ 
2021-05-01  & 59336.41& 0.62& 1.54e-12  &  3.08e-12 & 1.23e-13 &   5.51e-13     & 146.9&     890.0& 0.046\\ 
2021-07-17  & 59413.34& 0.75& 9.3e-13  &  1.45e-12 &    7e-14 &   3.41e-13     & 158.9&     480.0& 0.029\\ 
2021-08-17  & 59444.22& 0.8& 7e-13  &  1.08e-12 &  7.1e-14 &   2.46e-13     & 156.9&     360.0& 0.026\\ 
2021-08-24  & 59451.23& 0.81& 6.6e-13  &   9.7e-13 &  8.3e-14 &   2.28e-13     & 157.4&     340.0& 0.025\\ 
2021-09-01  & 59459.26& 0.83& 6.1e-13  &   9.1e-13 &  7.4e-14 &   2.37e-13     & 157.7&     320.0& 0.024\\ 
2021-10-13  & 59501.08& 0.9& 4.7e-13  &   7.2e-13 &  5.6e-14 &   3.24e-13     & 156.3&     260.0& 0.022\\ 
2021-10-22  & 59510.11& 0.91& 5.1e-13  &     7e-13 &  4.2e-14 &    3.5e-13     & 164.2&     270.0& 0.02\\ 
2022-02-11  & 59621.52& 1.1& 4.8e-13  &     6e-13 &  7.9e-14 &   2.22e-13     & 161.7&     240.0& 0.02\\ 
2022-03-08  & 59647.44& 1.15& 5.6e-13  &   7.9e-13 &  1.8e-14 &   1.13e-13     & 167.5&     270.0& 0.019\\ 
2022-03-21  & 59660.45& 1.17& 5.7e-13  &   9.4e-13 &    3e-14 &   1.01e-13     & 157.1&     290.0& 0.023\\ 
2022-05-09  & 59709.38& 1.25& 6e-13  &  1.84e-12 &  8.3e-14 &   3.34e-13     & 128.2&     460.0& 0.043\\ 
2022-06-11  & 59742.3& 1.31& 6.5e-13  &  2.68e-12 &  5.3e-14 &   4.38e-13     & 119.7&     590.0& 0.057\\ 
2022-08-30  & 59822.28& 1.44& 1.28e-12  &  3.38e-12 &  8.6e-14 &   5.05e-13     & 135.9&     860.0& 0.053\\ 
2022-10-01  & 59854.13& 1.49& 1.13e-12  &  3.35e-12 & 1.16e-13 &    5.6e-13     & 130.3&     830.0& 0.056\\ 
2023-03-02  & 60006.39& 1.75& 6.3e-13  &  1.03e-12 &  4.1e-14 &   1.78e-13     & 157.2&     330.0& 0.024\\ 
2023-06-10  & 60106.31& 1.92& 4.8e-13  &   6.3e-13 &  6.5e-14 &   2.92e-13     & 161.9&     250.0& 0.02\\ 
2023-08-07  & 60164.17& 2.02& 4.6e-13  &   6.2e-13 &  4.5e-14 &   2.98e-13     & 164.2&     240.0& 0.019\\

		\hline
	\end{tabular}
\end{table*}

The temperature of the hot component in the symbiotic star is estimated using an analytical relation derived by \cite{1981ASIC...69..517I}. This method relies on a case B recombination scenario, where the hot source ionizes the surrounding nebula. The relation considers the fluxes of H$\beta$, He I, and He II lines to determine the temperature, as shown in Equation (\ref{eq:ijima}). The temperature estimates are valid between 70,000 K and 200,000 K.

\begin{equation}
\label{eq:ijima}
T_{\rm{hot}} (\rm{in\,10^4\,K}) = 19.38 \sqrt{2.22F_{\rm{He \ II \ 4686}} \over4.16F_{\rm{H\,\beta}}+
9.94F_{\rm{He \ I \ 4471}}} + 5.13, 
\end{equation}

The luminosity of the hot component is determined by applying equation (8) of \cite{1991AJ....101..637K} and equation (6) of \cite{1997A&A...327..191M}. These calculations gave consistent results, with both estimates agreeing within 25 percent. The average of these estimates is provided in Table \ref{tab:hot}.
In addition, we estimated the radius by assuming a blackbody model for the hot component.

The estimated temperatures are given in Table \ref{tab:hot} and shown in Figure \ref{line:param}. Interestingly, temperatures calculated from emission line ratios show periodic behavior. This is entirely due to the geometrical effect caused by orbital motion. Due to the high angle of inclination of YY Her, the nebula surrounding the hot component is eclipsed by the giant star to varying degrees in different phases. The change in visibility of the nebular region leads to variations in line ratios, affecting the temperature estimation. It should be noted that the neutral lines (e.g., Balmer lines) originate from a wider area from the hot component compared to the high ionization lines (e.g., He II 4686 \AA{}), and show a higher degree of variation. Similar effects are observed in luminosity estimation, which is significantly influenced by the H$\beta$ line flux.  

Temperatures closer to \( \phi = 0.0 \) are overestimated, even exceeding values during the outburst (\( \geq1.4 \times 10^5 \) K).
Similar phase dependence of the estimated parameters has been reported by \cite{Combination_nova_sokoloski_2006} in Z And. We tried to address phase variation in temperature and luminosity calculations by introducing correction factors through fitting a sinusoidal function. However, we were unable to achieve a satisfactory fit. A periodic function to correct for the variation may not be sufficient, as this will not account for geometric effects that vary differently across different spectral lines.

The temperature and luminosity estimates obtained during the secondary eclipse are considered more reliable since the WD and the nebular region are directly exposed to the observer. Hence, we only considered the parameter estimated during secondary minima. During the outburst at (\( \phi = 0.55 \)), we estimated T = $\sim$1.41 x 10\textsuperscript{5} K, L = $\sim$ 1020 $L_{\odot}$, and R = 0.054 $R_{\sun}$. The values after one orbital cycle at (\( \phi = 1.49 \)) estimated as T = $\sim$1.30 x 10\textsuperscript{5} K, L = $\sim$ 830 $L_{\odot}$, and R = 0.056 $R_{\sun}$.

\section{Outburst Behavior}

The classical symbiotic star YY Her showed multiple outburst events in the past. A detailed follow-up study of the 1994-1996 outburst by \cite{1997ARep...41..802M} and \cite{2000ARep...44..190T} showed that the temperature of the hot component drops during the optical maxima. This indicates that the 1993 outburst was cool-type in nature. Based on line fluxes reported in Table 2 of  \cite{2000ARep...44..190T},  He II 4876 \AA{} $/$ H$\beta$  ratio during outburst (25 August 1993) was significantly lower than observation taken in a similar phase (\( \phi \sim 0.43 \)) two orbital cycles prior (09 May 1990), which is also in agreement with a cool type outburst.

In the current outburst, during the optical maximum, the temperature reaches approximately $\sim$1.4 x 10\textsuperscript{5} K, and the luminosity reaches $\sim$ 1020 $L_{\odot}$ compared to the same phase observation taken after 1 orbital cycle of YY Her, which yielded a temperature reaching approximately $\sim$ 1.3 x 10\textsuperscript{5} K, and the luminosity reaches $\sim$ 830 $L_{\odot}$. This confirms that the 2021 outburst of YY Her is of the hot type. Furthermore, the emission line ratios of He II to H$\beta$ were enhanced during the outburst, as estimated in this work and by \cite{2021ATel14464....1M}. Thus, we conclude that YY Her is showing both cool- and hot-type outbursts such as AG Dra.  
Earlier studies \cite{1997A&A...323..113M, 2013ATel.4996....1M} show that there were 4 outburst and brightening events and 8 smaller brightening events between 1890 and 2020. The major events have shown an amplitude $\geq$ 2 mag whereas the minor events are $\sim$ 1 mag. This is similar to what was reported in AG Dra, the smaller events are hot outbursts whereas the larger magnitude could be both cool or hot type outbursts. Compared to the major outburst reported in 1994, the magnitude of the 2021 outburst is much smaller ($\sim$ 1 mag), which is consistent with the hot type outburst.

\section{Summary}

We have presented a phase-resolved spectroscopic and photometric follow-up of YY Her over 1.5 orbital cycles after its 2021 outburst, combining HFOSC/HCT data with archival light curves.

Our analysis shows that the temperature and luminosity estimation of YY Her is affected by the orbital motion of the system, where the secondary contributes to the modulation of line fluxes, especially low-ionization lines. This causes overestimation of temperatures during the phases where the cool giant significantly covers the nebular region. Therefore, we consider the estimation of temperature and luminosity in phase 0.5, where the giant does not cover the nebular region and the WD, ensuring reliability.

During outburst at \( \phi = 0.55 \), the temperature reached $1.41 \times 10^5$ K, and the luminosity reached 1020 L\textsubscript{\(\odot\)} in the system. During \( \phi = 1.49 \) after the outburst, the temperature dropped approximately to $1.3 \times 10^5$ K, and the luminosity dropped approximately to 830 L\textsubscript{\(\odot\)} in the system. The post-outburst temperature and luminosity values are notably lower compared to those of the outburst (\( \phi = 0.55 \)). This is in agreement with a hot-type outburst. 

During the maxima of 1994-1996 outburst, YY Her showed a drop in temperature and He II 4876 \AA{} $/$ H$\beta$ ratio, which indicates a cool type outburst characteristics. 2013 $\&$ 2021 outbursts show the hot-type outburst characteristics. Hence, we conclude YY Her exhibits both cool and hot-type outbursts similar to AG Dra.

Ca II absorption lines at 8542\AA{} $\&$ 8662\AA{}  show a periodic variation half that of the orbital period. Since they originate from the cool giant, this is indirect evidence of the ellipsoidal effect. The observed secondary minima of YY Her originate from a combination of sinusoidal variation of the nebular contribution and ellipsoidal modulation of the distorted giant \citep{2002A&A...392..197M}. The appearance of a magnitude increase during secondary minima after the outburst is attributed to a heated WD at the line of sight (\( \phi = 0.5 \)), suggesting that the hot component may contribute to deeper minima in quiescence. Hence, one cannot rule out the possibility of contribution from eclipsing to the light curve. It could be possible that both effects may have a role in the observed light curve of the system.

%% The "ht!" tells LaTeX to put the figure "here" first, at the "top" next
%% and to override the normal way of calculating a float position

%% IMPORTANT! The old "\acknowledgment" command has be depreciated. It was
%% not robust enough to handle our new dual anonymous review requirements and
%% thus been replaced with the acknowledgment environment. If you try to 
%% compile with \acknowledgment you will get an error print to the screen
%% and in the compiled pdf.
%% 
%% Also note that the akcnowlodgment environment does not support long amounts of text. If you have a lot of people and institutions to acknowledge, do not use this command. Instead, create a new \section{Acknowledgments}.
%\begin{acknowledgments} 
\section*{Acknowledgments}

The authors thank the staff at IAO, Hanle, and Centre For Research \& Education in Science \& Technology (CREST), Hosakote, who helped carry out these observations. The IAO and CREST facilities are operated by the Indian Institute of Astrophysics, Bangalore.  We also thank all Himalayan Chandra Telescope (HCT) observers for accommodating dedicated time for Target of Opportunity (ToO) observations. We also express our gratitude to the HCT Time Allocation Committee (HTAC) for their support and dedication of time for both ToO and regular observations. This work made use of data from the GROWTH-India Telescope (GIT) set up by the Indian Institute of Astrophysics (IIA) and the Indian Institute of Technology Bombay (IITB) with funding from Indo-US Science and Technology Forum (IUSSTF) and Science and Engineering Research Board, Department of Science and Technology, Government of India (DST-SERB). It is located at IAO. We acknowledge funding by the IITB alumni batch of 1994, which partially supports the operations of the telescope. We thank the anonymous reviewer for comments and suggestions.  
This study used the open source package Astropy,\footnote{http://www.astropy.org}, a Python package for Astronomy \citep{2018AJ....156..123A}. We acknowledge use of Zwicky Transient Facility data obtained via the NASA/IPAC Infrared Science Archive (IRSA) ZTF Image Service.
%\end{acknowledgments}

%% To help institutions obtain information on the effectiveness of their 
%% telescopes the AAS Journals has created a group of keywords for telescope 
%% facilities.
%
%% Following the acknowledgments section, use the following syntax and the
%% \facility{} or \facilities{} macros to list the keywords of facilities used 
%% in the research for the paper. Each keyword is check against the master 
%% list during copy editing. Individual instruments can be provided in 
%% parentheses, after the keyword, but they are not verified.

\vspace{5mm}
\facilities{HCT, GIT, ASAS-SN, PO:1.2m (ZTF), AAVSO, Gaia}

%% Similar to \facility{}, there is the optional \software command to allow 
%% authors a place to specify which programs were used during the creation of 
%% the manuscript. Authors should list each code and include either a
%% citation or url to the code inside ()s when available.

\software{astropy \citep{2013A&A...558A..33A,2018AJ....156..123A};
          IRAF \citep{1993ASPC...52..173T};
            PYRAF\citep{2012ascl.soft07011S}}

%% Appendix material should be preceded with a single \appendix command.
%% There should be a \section command for each appendix. Mark appendix
%% subsections with the same markup you use in the main body of the paper.

%% Each Appendix (indicated with \section) will be lettered A, B, C, etc.
%% The equation counter will reset when it encounters the \appendix
%% command and will number appendix equations (A1), (A2), etc. The
%% Figure and Table counter will not reset.

\appendix
\restartappendixnumbering

\section{Supplementary Data and Figures}

\begin{figure*}
\centering
\includegraphics[width=\columnwidth]{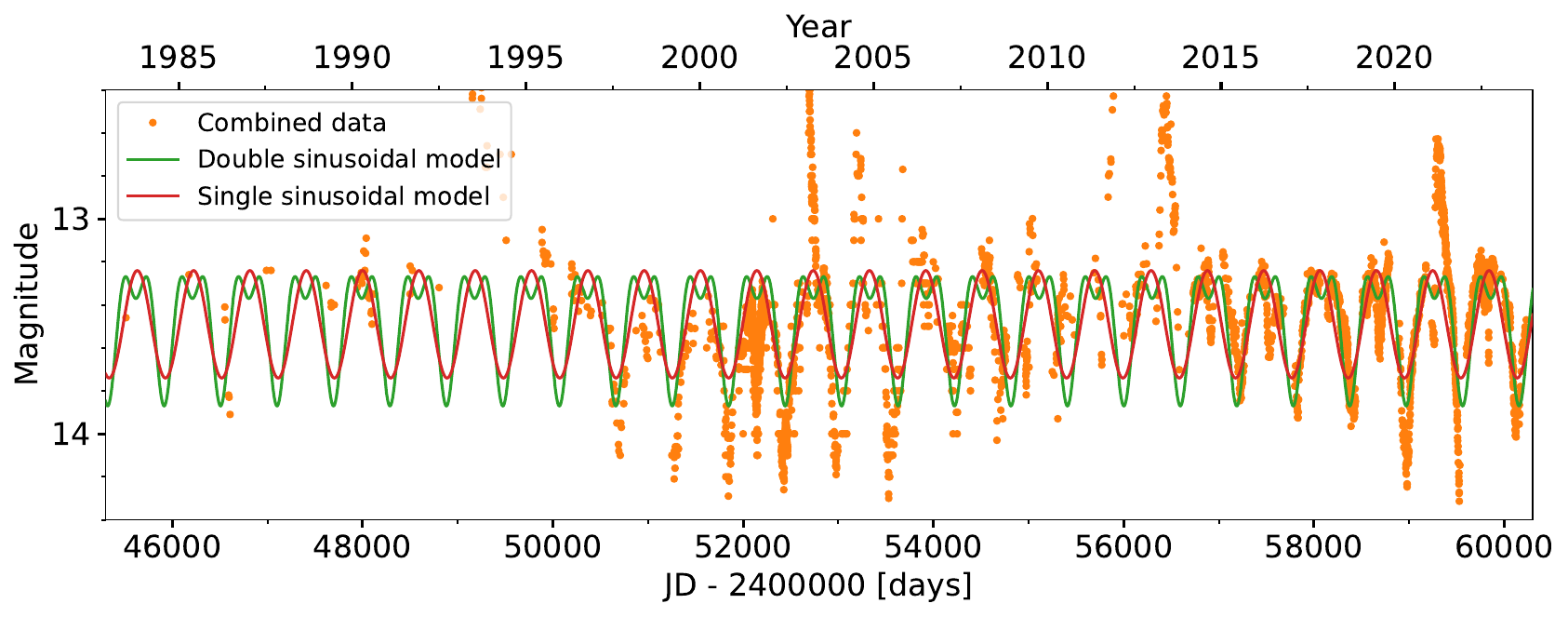}
	\caption{Single and double sinusoidal models fit on the combined V and g band data. Ephemeris given by Equation \ref{eq:ephemeris} is adopted from double sinusoidal model.}
	\label{plot:double}
\end{figure*}

\begin{figure}
\begin{center}

    \includegraphics[width=\columnwidth]{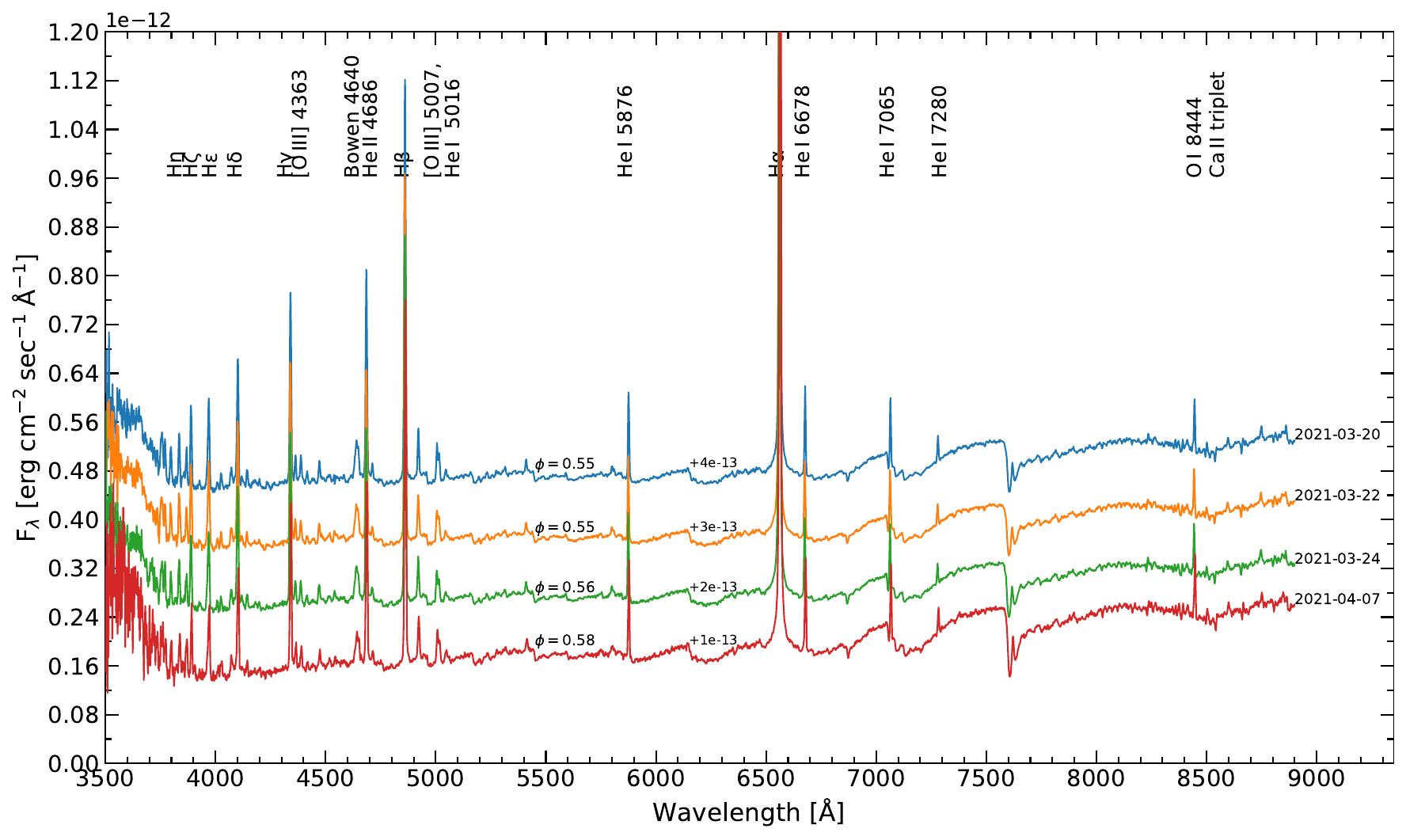}
	\caption{Optical low‐resolution spectra of \object{YY~Her} (not shown in Fig. \ref{spectra1}), plotted in chronological order. The identified lines are labeled on the top. Observation date and orbital phase ($\phi$) are given for each spectrum. Spectra are shifted to the indicated amount for better visibility.}
    \label{spectra2}
\end{center}
\end{figure}

The Appendix contains GIT griz photometry (Table \ref{git_phot}), dereddened line fluxes from HCT spectra (Table \ref{flux_table}), single- and double-sinusoidal fits to V- and g-band photometry (Figure \ref{plot:double}), and supplementary spectra (Figure \ref{spectra2}).

\begin{deluxetable}{ccccccccc}
\tablecaption{GIT photometric data \label{git_phot}}
\tablewidth{0pt}
\tablehead{
\colhead{JD - 2400000} & \multicolumn{2}{c}{GIT g} & \multicolumn{2}{c}{GIT i} & \multicolumn{2}{c}{GIT r} & \multicolumn{2}{c}{GIT z} \\
\colhead{} & \colhead{Mag} & \colhead{Error} & \colhead{Mag} & \colhead{Error} & \colhead{Mag} & \colhead{Error} & \colhead{Mag} & \colhead{Error}
}
\startdata
    59618.5 &           13.510 &                  0.044 &           12.300 &                  0.023 &           11.226 &                  0.020 &           10.640 &                  0.052 \\
    59626.5 &           13.340 &                  0.049 &           12.108 &                  0.049 &               -- &                     -- &               -- &                     -- \\
    59631.4 &           13.424 &                  0.027 &           12.136 &                  0.045 &           11.205 &                  0.029 &               -- &                     -- \\
    59631.5 &               -- &                     -- &               -- &                     -- &               -- &                     -- &           10.592 &                  0.054 \\
    59636.4 &               -- &                     -- &               -- &                     -- &               -- &                     -- &           10.557 &                  0.060 \\
    59644.4 &           13.522 &                  0.064 &           12.299 &                  0.049 &           11.268 &                  0.074 &           10.635 &                  0.059 \\
    59649.4 &           13.452 &                  0.047 &           12.229 &                  0.040 &           11.189 &                  0.049 &           10.617 &                  0.060 \\
    59659.4 &           13.388 &                  0.021 &           12.090 &                  0.047 &           11.181 &                  0.039 &           10.640 &                  0.067 \\
    59664.4 &           13.363 &                  0.042 &           12.154 &                  0.061 &           11.164 &                  0.082 &           10.623 &                  0.068 \\
    59675.3 &           13.293 &                  0.070 &           12.417 &                  0.088 &           11.115 &                  0.090 &           10.597 &                  0.109 \\
    59680.3 &           13.307 &                  0.037 &           12.194 &                  0.049 &               -- &                     -- &           10.658 &                  0.061 \\
    59700.3 &           13.175 &                  0.054 &           12.136 &                  0.052 &               -- &                     -- &           10.542 &                  0.056 \\
    59716.2 &           13.288 &                  0.103 &           11.984 &                  0.050 &           11.169 &                  0.084 &           10.583 &                  0.074 \\
    59722.2 &           13.166 &                  0.059 &           12.226 &                  0.043 &           11.072 &                  0.045 &           10.677 &                  0.093 \\
    59722.3 &               -- &                     -- &           12.080 &                  0.059 &           11.186 &                  0.057 &           10.735 &                  0.092 \\
    59722.4 &           13.285 &                  0.072 &               -- &                     -- &               -- &                     -- &               -- &                     -- \\
    59728.2 &           13.212 &                  0.084 &           11.891 &                  0.057 &           11.052 &                  0.033 &           10.551 &                  0.053 \\
    59737.1 &               -- &                     -- &               -- &                     -- &               -- &                     -- &           10.549 &                  0.079 \\
    59737.2 &           13.218 &                  0.103 &           11.935 &                  0.043 &           11.100 &                  0.031 &               -- &                     -- \\
    59738.4 &           13.305 &                  0.102 &           11.989 &                  0.050 &           11.167 &                  0.040 &           10.624 &                  0.042 \\
    59741.4 &           13.102 &                  0.169 &               -- &                     -- &               -- &                     -- &               -- &                     -- \\
    59743.1 &           13.226 &                  0.090 &           11.974 &                  0.031 &           11.104 &                  0.087 &           10.565 &                  0.124 \\
    59749.2 &               -- &                     -- &               -- &                     -- &           11.132 &                  0.030 &               -- &                     -- \\
    59749.3 &               -- &                     -- &               -- &                     -- &               -- &                     -- &           10.628 &                  0.079 \\
    59753.3 &               -- &                     -- &               -- &                     -- &               -- &                     -- &           10.607 &                  0.051 \\
    59763.2 &           13.194 &                  0.048 &           11.952 &                  0.046 &           11.085 &                  0.036 &           10.598 &                  0.046 \\
    59769.4 &               -- &                     -- &           11.989 &                  0.074 &               -- &                     -- &           10.584 &                  0.136 \\
    59785.3 &               -- &                     -- &           12.090 &                  0.027 &           11.258 &                  0.036 &           10.736 &                  0.049 \\
    59820.1 &               -- &                     -- &           12.243 &                  0.052 &           11.279 &                  0.028 &           10.864 &                  0.108 \\
    59820.2 &           13.145 &                  0.030 &               -- &                     -- &               -- &                     -- &               -- &                     -- \\
    59850.2 &           13.307 &                  0.071 &               -- &                     -- &           11.436 &                  0.134 &           11.102 &                  0.162 \\
    59853.2 &           13.273 &                  0.043 &           12.175 &                  0.031 &           11.403 &                  0.033 &           10.858 &                  0.045 \\
    59855.1 &           13.221 &                  0.061 &           12.095 &                  0.059 &           11.331 &                  0.062 &           10.747 &                  0.075 \\
    59858.2 &           13.215 &                  0.055 &           12.058 &                  0.056 &               -- &                     -- &           10.680 &                  0.097 \\
    59873.1 &           13.222 &                  0.048 &               -- &                     -- &           11.256 &                  0.058 &               -- &                     -- \\
    59882.1 &           13.190 &                  0.026 &           12.022 &                  0.040 &           11.246 &                  0.049 &           10.767 &                  0.082 \\
    59894.1 &               -- &                     -- &               -- &                     -- &               -- &                     -- &           10.661 &                  0.073 \\
    60070.2 &               -- &                     -- &               -- &                     -- &           11.306 &                  0.050 &               -- &                     -- \\
    60070.3 &           13.584 &                  0.030 &           12.386 &                  0.023 &               -- &                     -- &           10.687 &                  0.050 \\
    60075.2 &           13.647 &                  0.032 &           12.287 &                  0.085 &           11.311 &                  0.027 &               -- &                     -- \\
    60085.2 &           13.653 &                  0.030 &           12.502 &                  0.035 &           11.513 &                  0.034 &           10.780 &                  0.067 \\
    60090.3 &               -- &                     -- &           12.488 &                  0.030 &           11.385 &                  0.043 &           10.708 &                  0.071 \\
    60090.4 &           13.673 &                  0.024 &               -- &                     -- &               -- &                     -- &               -- &                     -- \\
    60101.2 &               -- &                     -- &           12.665 &                  0.037 &               -- &                     -- &               -- &                     -- \\
    60106.1 &           13.972 &                  0.039 &           12.721 &                  0.035 &           11.486 &                  0.031 &               -- &                     -- \\
    60106.2 &               -- &                     -- &               -- &                     -- &               -- &                     -- &           10.799 &                  0.042 \\
    60111.2 &           13.966 &                  0.071 &           12.675 &                  0.040 &               -- &                     -- &               -- &                     -- \\
    60116.3 &           13.951 &                  0.031 &           12.720 &                  0.057 &           11.652 &                  0.049 &           10.801 &                  0.085 \\
\enddata
\tablecomments{
Photometric data from GIT observations in g, r, i, and z filters. Magnitudes and associated errors are provided; Julian Dates are rounded to the nearest 0.1 day.
}
\end{deluxetable}

\begin{splitdeluxetable*}{cccccccccccccBcccccccc}
\tabletypesize{\footnotesize}
\tablewidth{0pt}
\tablecaption{De-reddened line flux measurements}
\label{flux_table}
\tablehead{
\colhead{JD - 2400000} & 
\colhead{H$\alpha$} & 
\colhead{Error} & 
\colhead{H$\beta$} & 
\colhead{Error} & 
\colhead{H$\gamma$} & 
\colhead{Error} & 
\colhead{H$\delta$} & 
\colhead{Error} & 
\colhead{He II 4686} & 
\colhead{Error} & 
\colhead{He I 4471} & 
\colhead{Error} & 
\colhead{He I 5876} & 
\colhead{Error} & 
\colhead{He I 6678} & 
\colhead{Error} & 
\colhead{Ca II 8542} & 
\colhead{Error} &
\colhead{Ca II 8662} &
\colhead{Error}
}
\startdata
59292.481 &   998.0 &        11.0 &  370.1 &        4.7 &   152.1 &         5.4 &   119.8 &         3.3 &      162.7 &            3.3 &     14.50 &          1.30 &      71.8 &           1.6 &      74.9 &           1.7 &        9.9 &            1.8 &        5.4 &            1.2 \\
59294.478 &  1140.0 &        11.0 &  369.6 &        4.5 &   162.3 &         5.8 &   114.2 &         3.6 &      180.1 &            3.6 &     18.30 &          1.80 &      72.4 &           1.7 &      73.6 &           1.6 &       10.3 &            1.6 &        7.5 &            1.6 \\
59296.437 &  1088.0 &        12.0 &  367.9 &        4.3 &   156.2 &         4.9 &   116.0 &         2.9 &      162.6 &            2.9 &     18.60 &          1.40 &      73.0 &           1.7 &      68.9 &           1.7 &        8.9 &            1.5 &        6.8 &            2.0 \\
59298.411 &  1093.0 &        14.0 &  364.6 &        4.7 &   152.4 &         5.0 &   103.7 &         3.1 &      162.1 &            3.2 &     16.00 &          1.40 &      73.4 &           1.9 &      68.9 &           1.8 &       10.0 &            1.4 &        8.2 &            2.9 \\
59300.424 &  1095.0 &        14.0 &  364.5 &        4.8 &   155.0 &         5.3 &   112.1 &         3.0 &      165.5 &            3.8 &     17.20 &          1.50 &      69.3 &           1.8 &      67.7 &           1.8 &        8.8 &            1.5 &        6.9 &            1.9 \\
59308.479 &  1267.0 &        13.0 &  380.4 &        4.4 &   150.1 &         5.3 &   113.3 &         2.8 &      170.0 &            3.8 &     16.40 &          1.40 &      71.8 &           1.7 &      77.0 &           1.7 &       13.5 &            1.7 &        8.4 &            2.1 \\
59312.338 &  1384.0 &        15.0 &  383.6 &        5.0 &   135.6 &         5.5 &    97.5 &         3.2 &      165.4 &            4.2 &     19.70 &          2.40 &      71.3 &           2.2 &      83.6 &           2.1 &       12.8 &            2.3 &        4.7 &            2.0 \\
59330.402 &  1014.3 &         9.1 &  339.6 &        3.6 &   139.9 &         5.9 &   110.5 &         2.9 &      173.5 &            3.1 &     15.70 &          1.50 &      65.3 &           1.4 &      67.4 &           1.3 &       15.6 &            1.4 &       11.1 &            2.2 \\
59336.407 &   827.0 &        15.0 &  308.1 &        5.5 &   122.9 &         5.3 &    87.0 &         2.5 &      153.9 &            3.6 &     12.30 &          1.30 &      55.1 &           2.0 &      54.0 &           2.0 &       13.7 &            1.9 &       11.6 &            2.4 \\
59413.342 &   620.0 &        14.0 &  145.0 &        3.3 &    53.0 &         2.5 &    32.1 &         1.5 &       93.4 &            2.7 &      7.00 &          1.10 &      34.1 &           1.6 &      37.6 &           1.9 &       19.6 &            3.9 &       14.7 &            2.6 \\
59444.219 &   509.0 &        15.0 &  108.2 &        3.5 &    35.3 &         1.9 &    24.6 &         1.3 &       69.7 &            2.4 &      7.10 &          1.10 &      24.6 &           1.5 &      26.3 &           2.0 &       22.4 &            6.5 &       14.4 &            3.7 \\
59451.235 &   455.0 &        18.0 &   96.9 &        3.9 &    33.2 &         2.5 &    19.9 &         1.6 &       65.6 &            2.8 &      8.30 &          1.40 &      22.8 &           1.9 &      24.0 &           2.4 &       25.6 &            9.7 &       15.6 &            6.6 \\
59459.259 &   429.0 &        14.0 &   90.8 &        3.3 &    31.1 &         2.2 &    19.3 &         1.6 &       61.2 &            2.7 &      7.40 &          1.20 &      23.7 &           1.7 &      19.6 &           2.1 &       23.0 &            4.7 &       17.5 &            3.5 \\
59501.079 &   411.0 &        16.0 &   72.3 &        3.6 &    23.0 &         2.6 &    15.4 &         1.4 &       47.1 &            2.5 &      5.60 &          1.00 &      32.4 &           2.1 &      13.6 &           1.7 &       19.4 &            6.9 &       14.7 &            3.9 \\
59510.113 &   406.0 &        19.0 &   70.4 &        3.6 &    22.6 &         2.7 &    16.1 &         1.9 &       51.2 &            3.1 &      4.20 &          1.10 &      35.0 &           2.3 &      12.7 &           1.9 &       15.3 &            3.6 &       14.3 &            4.5 \\
59621.515 &   377.0 &        12.0 &   60.4 &        3.0 &    17.1 &         2.9 &    25.0 &         2.9 &       48.2 &            2.4 &      7.90 &          1.60 &      22.2 &           1.8 &      19.0 &           2.2 &       30.2 &            4.2 &       24.4 &            4.5 \\
59647.436 &   390.0 &        11.0 &   78.8 &        2.5 &    28.1 &         2.4 &    23.1 &         1.3 &       56.0 &            2.1 &      1.82 &          0.77 &      11.3 &           1.3 &      17.8 &           1.7 &       28.5 &            3.1 &       18.5 &            3.5 \\
59660.448 &   422.0 &        15.0 &   94.0 &        4.0 &    32.9 &         2.8 &    22.4 &         1.5 &       56.6 &            2.7 &      3.00 &          1.10 &      10.1 &           1.7 &      20.1 &           2.4 &       25.6 &            3.9 &       17.9 &            3.9 \\
59709.379 &   693.0 &        16.0 &  184.4 &        4.1 &    75.0 &         2.8 &    44.9 &         1.6 &       60.3 &            2.2 &      8.30 &          1.10 &      33.4 &           1.7 &      37.8 &           2.0 &       17.1 &            3.1 &       11.9 &            3.2 \\
59742.300 &   883.0 &        15.0 &  267.8 &        4.9 &   101.3 &         3.0 &    61.6 &         2.1 &       65.4 &            2.3 &      5.30 &          1.00 &      43.8 &           1.6 &      48.8 &           1.8 &       15.6 &            5.2 &       12.5 &            3.6 \\
59822.277 &   785.0 &        27.0 &  337.6 &        9.5 &   135.1 &         5.2 &    83.1 &         3.1 &      127.7 &            4.0 &      8.60 &          1.30 &      50.5 &           2.5 &      61.6 &           3.0 &       15.7 &            7.0 &        7.8 &            4.9 \\
59854.130 &  1209.0 &        34.0 &  335.1 &        7.3 &   142.4 &         4.4 &    91.2 &         2.6 &      112.9 &            3.9 &     11.60 &          1.10 &      56.0 &           2.3 &      78.6 &           2.8 &       12.2 &            8.3 &        6.5 &            2.9 \\
60006.391 &   482.0 &        21.0 &  103.3 &        5.0 &    37.1 &         2.7 &    23.4 &         1.7 &       63.3 &            3.5 &      4.10 &          1.40 &      17.8 &           2.3 &      19.4 &           3.1 &       25.8 &            5.2 &       20.2 &            6.4 \\
60106.309 &   365.0 &        17.0 &   63.5 &        3.4 &    21.3 &         2.7 &    32.5 &         3.6 &       48.3 &            3.0 &      6.50 &          1.30 &      29.2 &           3.3 &      10.6 &           2.5 &       14.7 &            4.2 &       17.2 &            7.0 \\
60164.172 &   343.0 &        24.0 &   62.2 &        5.0 &    17.6 &         2.9 &    18.4 &         2.2 &       46.3 &            3.3 &      4.50 &          2.10 &      29.8 &           2.8 &      14.2 &           2.9 &       23.0 &           27.0 &       16.0 &           11.0 \\
\enddata
\tablecomments{All flux values are in units of $\times$ $10^{-14}$ erg cm$^{-2}$ s$^{-1}$. The uncertainties represent propagated errors. Ca II lines are absorption lines, whereas all other lines are in emission.}
\end{splitdeluxetable*}

%% For this sample we use BibTeX plus aasjournals.bst to generate the
%% the bibliography. The sample631.bib file was populated from ADS. To
%% get the citations to show in the compiled file do the following:
%%
%% pdflatex sample631.tex
%% bibtext sample631
%% pdflatex sample631.tex
%% pdflatex sample631.tex

\bibliography{yyher}
\bibliographystyle{aasjournal}

%% This command is needed to show the entire author+affiliation list when
%% the collaboration and author truncation commands are used. It has to
%% go at the end of the manuscript.
%\allauthors

%% Include this line if you are using the \added, \replaced, \deleted
%% commands to see a summary list of all changes at the end of the article.
%\listofchanges

\end{document}